\documentclass[lettersize,journal]{IEEEtran}
\pdfoutput=1
\UseRawInputEncoding
\usepackage{amsmath,amsfonts}
\usepackage{algorithmicx}
\usepackage[ruled]{algorithm2e}
\usepackage{array}
\usepackage[caption=false,font=normalsize,labelfont=sf,textfont=sf]{subfig}
\usepackage{textcomp}
\usepackage{stfloats}
\usepackage{url}
\usepackage{verbatim}
\usepackage{graphicx}
\usepackage{cite}
\usepackage{diagbox}
\usepackage{cases}
\usepackage{bm}
\usepackage{amssymb}
\usepackage{multirow}
\usepackage{epstopdf}
\usepackage{epsfig}
 \usepackage{amsthm}
 \usepackage{color}
 \usepackage{booktabs}
\newtheorem{theorem}{Theorem}
\newtheorem{lemma}{Lemma}
\newtheorem{remark}{Remark}
\begin{document}

\title{LRIP-Net: Low-Resolution Image Prior based Network for Limited-Angle CT Reconstruction}

\author{Qifeng Gao,
	Rui Ding,
	Linyuan Wang,
	Bin Xue,
	and Yuping Duan$^*$
\thanks{This work did not involve human subjects or
animals in its research. All authors declare that they have no known conflicts of interest in terms of competing financial interests or personal relationships that could have an influence or are relevant to the work reported in this paper.
The work was partially supported by the National Natural Science Foundation of China (NSFC 12071345, 11701418), Major Science and Technology Project of Tianjin 18ZXRHSY00160, and Recruitment Program of Global Young Expert. The fourth author B Xue was supported by the National Natural Science Foundation of China (NSFC 62075162,  62001329) and Natural Science Foundation of Tianjin 19JCQNJC01700. \emph{Asterisk indicates the corresponding author.}}
\thanks{Qifeng Gao is with the Center for Applied Mathematics, Tianjin University,
	Tianjin 300072, China (e-mail: gaoqifeng\_98@tju.edu.cn).}
\thanks{Rui Ding is with the Center for Applied Mathematics, Tianjin University,
	Tianjin 300072, China (e-mail: d18822058320@163.com).}
\thanks{Linyuan Wang is with Henan Key Laboratory of Imaging and Intelligent Processing, PLA Strategy Support Force Information Engineering University, Zhengzhou 450002, China (e-mail:wanglinyuanwly@163.com).}
\thanks{Bin Xue is with the School of Marine Science and Technology, TianJin University, Tianjin 30072, China (e-mail: xuebin@tju.edu.cn).}
\thanks{Yuping Duan is with the Center for Applied Mathematics, Tianjin University, Tianjin 300072, China (e-mail: yuping.duan@tju.edu.cn).}
}

\markboth{Journal of \LaTeX\ Class Files,~Vol.~14, No.~8, August~2021}%
{Shell \MakeLowercase{\textit{et al.}}: A Sample Article Using IEEEtran.cls for IEEE Journals}


\maketitle

\begin{abstract}
In the practical applications of computed tomography imaging, the projection data may be acquired within a limited-angle range and corrupted by noises due to the limitation of scanning conditions. The noisy incomplete projection data results in the ill-posedness of the inverse problems. In this work, we theoretically verify that the low-resolution reconstruction problem has better numerical stability than the high-resolution problem. In what follows, a novel low-resolution image prior based CT reconstruction model is proposed to make use of the low-resolution image to improve the reconstruction quality. More specifically, we build up a low-resolution reconstruction problem on the down-sampled projection data, and use the reconstructed low-resolution image as prior knowledge for the original limited-angle CT problem. We solve the constrained minimization problem by the alternating direction method with all subproblems approximated by the convolutional neural networks. Numerical experiments demonstrate that our double-resolution network outperforms both the variational method and popular learning-based reconstruction methods on noisy limited-angle reconstruction problems.
\end{abstract}


\section{Introduction}
\IEEEPARstart{X}{-ray} Computed Tomography (CT) is widely used for clinical diagnosis, the quality of which directly affects the judgment of the clinicians. The filtered back-projection (FBP), algebraic reconstruction technique (ART), and simultaneous algebraic reconstruction technique (SART) are popular choices for the full scanned CT data. However, the imaging system may collect incomplete projection data due to the system's geometric limitations and other factors. It was proven that the reconstruction problem becomes highly unstable with scanning angular range less than $2\pi/3$. For more background, we refer to \cite{Frikel2013} and references therein. The aforementioned degenerated scanned data makes the direct method and iterative methods suffer from severe streaking artifacts and noise-induced artifacts \cite{He2019,Goy2019}. For limited-angle CT reconstruction, various algorithms were proposed, which can be broadly categorized into the regularization-based method and learning-based method.

The regularization-based method has been applied to the improperly posed problems and achieved great successes, which are good choices for the limited-angle CT reconstruction problems. Since compressed sensing (CS) was proposed by Candes, Romberg and Tao \cite{Candes2006}, various models and algorithms have been proposed to improve the reconstruction quality using the total variation (TV) for image reconstruction problems \cite{Lustig2008,Chen2008}. Sidky and Pan \cite{2008Image} developed the primal-dual algorithm for solving the TV minimization problem, which performed well concerning angular under-sampling reconstruction. Ritschl \emph{et al.} \cite{Ritschl2011} presented a new method for optimized parameter adaption for sparsity constrained image reconstruction. Chen \emph{et al.} \cite{chen2013limited} proposed the anisotropic TV minimization method, which performed better than the isotropic TV model for the limited-angle CT reconstruction.  Frikel \cite{Frikel2013} used the sparse regularization technique in combination with curvelets to realize an edge-preserving reconstruction. Cai \emph{et al.} \cite{2014Edge} developed an edge guided total variation minimization reconstruction algorithm in dealing with high quality image reconstruction. Xu \emph{et al.} \cite{Xu2019} combined the $\ell_1$ norm of gradient and $\ell_0$ norm of the gradient as the regularization term and developed the efficient alternating edge-preserving diffusion and smoothing algorithm, which can well preserve the edges for limited-angle reconstruction problem.

In addition to TV-based regularization methods, high-order regularization methods have also been investigated for degenerated scanned data. Niu \emph{et al.} \cite{2014Sparse} presented a penalized weighted least-squares scheme to retain the image quality by incorporating the total generalized variation regularization. Zhang \emph{et al.} \cite{2017Euler} introduced the curvature-driven Euler's elastica regularization to rectify large curvatures and kept the isophotes smooth without erratic distortions. Cai \emph{et al.} \cite{2017Block} proposed the block matching sparsity regularization for CT image reconstruction for an incomplete projection set. Wang \emph{et al.} \cite{wang2019guided} presented the guided image filtering-based limited-angle CT reconstruction algorithm using wavelet frame. Xu \emph{et al.} \cite{9084157} combined the dictionary learning and image gradient $\ell_0$-norm into image reconstruction model for limited-angle CT reconstruction. Wang \emph{et al.} \cite{Wang} considered minimizing the $\ell_1/\ell_2$ term on the gradient for a limited-angle scanning problem in CT reconstruction. However, the aforementioned regularization methods are usually time-consuming and suffer from tricky parameter tuning.

Due to the development of deep convolutional neural networks (CNNs) in a broad range of computer vision tasks, deep learning methods become more and more popular in the medical imaging field. With regard to limited-angle CT reconstruction, Pelt and Batenburg \cite{pelt2013fast} proposed an artificial neural network-based fast limited-angle image reconstruction algorithm, which can be regarded as a weighted combination of the FBP method and some learned filters. Boublil \emph{et al.} \cite{boublil2015spatially} utilized a CNN-based model to integrate multiple reconstructed results. Kang \emph{et al.} \cite{kang2017deep} constructed a deep CNN model in the wavelet domain, which trained the wavelet coefficients from the CT images after applying the contourlet transform. Gupta \emph{et al.} \cite{gupta2018cnn} presented a new image reconstruction method, which replaced the projector in a projected gradient descent with a CNN. Adler and {\"O}ktem \cite{adler2018learned} proposed a deep neural network by unrolling a proximal primal-dual optimization method and replacing the proximal operators with a CNN. Chen \emph{et al.} \cite{chen2018learn} unfolded the field of experts regularized CT reconstruction model into a deep learning network, all parameters of which can be learned from the training process. Han and Ye \cite{han2018framing} proposed a new multi-resolution deep learning scheme based on the frame condition to overcome the limitation of U-net. Zhang \emph{et al.} \cite{zhang2019jsr} presented a new deep CNN jointly reconstructs CT images and their associated Radon domain projections, and constructed a hybrid loss function to effectively protect the important structure of images. Bubba \emph{et al.} \cite{bubba2019learning} developed a hybrid reconstruction framework that fused model-based sparse regularization with data-driven deep learning to solve the severely ill-posed inverse problem of limited-angle CT.  Arridge  \emph{et al.} \cite{arridge2019solving} attempted to provide an overview of methods for integrating data-driven concepts into the field of inverse problems and a solid mathematical theory. W{\"u}rfl \emph{et al.} \cite{wurfl2018deep} mapped the Feldkamp-Davis-Kress algorithm to the neural networks by introducing a novel cone-beam back-projection layer for limited-angle problems. Lin \emph{et al.} \cite{lin2019dudonet} proposed an end-to-end trainable dual-domain network to simultaneously restore sinogram consistency and enhance CT images. Baguer \emph{et al.} \cite{baguer2020computed} introduced the deep image prior approach in combination with classical regularization and an initial reconstruction. Ding \emph{et al.} \cite{ding2020low} came up with a method based on the unrolling of a proximal forward-backward splitting framework with a data-driven image regularization via deep neural networks. Cheng \emph{et al.} \cite{cheng2020learned} proposed a novel reconstruction model to jointly reconstruct a high-quality image and its corresponding high-resolution projection data. Zang \emph{et al.} \cite{Zang_2021_ICCV} proposed IntraTomo, a powerful framework that combines the benefits of learning-based and model-based approaches for solving highly ill-posed inverse problems. Hu \emph{et al.} \cite{9496261} developed a method termed Single-shot Projection Error Correction Integrated Adversarial Learning (SPECIAL) progressive-improvement strategy, which could effectively combine the complementary information contained in the image domain and projection domain. Bubba \emph{et al.} \cite{doi:10.1137/20M1343075} proposed a novel CNN, designed for learning pseudodifferential operators in the context of linear inverse problems. Hu \emph{et al.} \cite{hu2022dior} proposed a novel reconstruction framework termed Deep Iterative Optimization-based Residual-learning (DIOR) for limited-angle CT, which combined iterative optimization and deep learning based on the residual domain.  Although the aforementioned learning-based methods have achieved better reconstruction results than the regularization-based methods, the high ill-posedness of the limited-angle reconstruction problems still challenges the reconstruction quality.

The incomplete projection data makes the inverse problem towards the ill-posedness, becoming more and more sensitive to noises. Through theoretical analysis, we find out that the low-resolution problem tends to be more numerically stable than the high-resolution problem supposing that the scanning range remains the same. Then, we propose a novel low-resolution image prior based trainable reconstruction approach for the limited-angle CT reconstruction. More specifically, we use the established reconstruction method to obtain the low-resolution image from the down-sampled raw measured data. In what follows, we build up the constrained reconstruction problem, which is solved by the alternating direction method. By approximating the resolvent operators by CNNs, an end-to-end algorithm is gained to reconstruct images from raw data. We evaluate the performance of the proposed method on the American
Association of Physicists in Medicine (AAPM) Challenge dataset. By comparing with the state-of-the-art reconstruction methods, our algorithm is shown more effective in dealing with limited-angle data contaminated by either Gaussian or Poisson noises.

The rest of the paper is organized as follows.
We present the double-resolution reconstruction method and its theoretical foundation in Section \ref{sec2}. Section \ref{sec3} is dedicated to explaining the details of our network architecture, loss function, and optimization. We present the numerical results on AAPM phantom CT dataset corrupted by Gaussian noises and Poisson noises in Section \ref{sec4}.
Finally, a brief conclusion and possible future works are presented in Section \ref{sec6}.

\section{The double-resolution reconstruction method}
\label{sec2}
\subsection{Double-resolution reconstruction model}
The CT reconstruction problem aims to reconstruct clean image $\bm u \in X$ from the projection data $\bm f \in Y$ with unknown noise $\bm \delta \in Y$, whose mathematical formulation is:
\begin{equation}\label{inverse problem}
	   \bm f=A \bm u+\bm \delta,
\end{equation}
where the reconstruction space $X$ and the data space $Y$ are typically Hilbert Spaces, $A : X \rightarrow Y$ is the forward operator that models how an image gives rise to data in absence of noise. When the light source is a fan beam, the dimension of the system matrix is $M\times N$ with $M$ and $N$ given as follows
\[M=N_{\rm{views}}\times N_{\rm{bins}},~~ \mbox{and}~~ N=n\times n,\]
where $N_{\rm{views}}$ denotes the number of angles in the angular interval for limited-angle CT reconstruction, $N_{\rm{bins}}$ denotes the number of units on the detector, and $N$ represents the number of pixels of the input image.
For CT reconstruction problem \eqref{inverse problem}, the condition number of the system matrix $A$ directly affects the stability of the solution \cite{Jorgensen2013}. The larger the condition number of the system matrix, the more serious the ill-posedness of the inverse problem, which may result in the degradation of the numerical methods.

Let us define a low-resolution system matrix $A_l$, the geometric parameters of which are consistent with the system matrix $A$, i.e., $N_{\rm{views}}$ and $N_{\rm{bins}}$ of $A_l$ being the same as $A$, and only the number of the pixels changes. More specifically, we use the equidistant sampling with the down-sampling rate $1/\tau$. Then the dimension of the low-resolution system matrix $A_l$ becomes $M\times N/\tau^2$, and the low-resolution image can be expressed by the down-sampling matrix $D$ as given below
\[		\bm{u}_l=D\bm u,  \]
where $\bm u_l$ is the down-sampled image and $D^TD$ is a diagonal matrix with diagonal elements being either 1 or 0.  Then we can arrive at the main theorem supporting our observation that a low-resolution problem has better numerical stability.

\begin{theorem}\label{theorem1}
Let $A$ and $A_l$ be the system matrix for the reconstruction problem $Au = f$ and $A_l u_l =f_l$, respectively, where $f_l = A_l DA^+ f$ with $D$ being a down-sampling operator.
Then in the sense of Moore-Penrose generalized inverse, we have
	\[cond(A_{l})\le cond(A).\]
\end{theorem}
\begin{IEEEproof}	
Since the down-sampling matrix $D$ is defined by the equidistant sampling, the diagonal elements of $D$ are either 0 or 1, and $D^TD$ is the diagonal matrix. Then we directly have
	\begin{equation*}
		\|D\|_{2}=\|D^{+}\|_{2}=1,
	\end{equation*}
where $D^{+}$ represents the the Moore-Penrose generalized inverse matrix of $D$, and $\|D\|_{2}=\sqrt{\lambda_{max}\left(D^TD\right)}$ with $\lambda_{max}$ representing the largest eigenvalue of $D^TD$. From given conditions, we have
\[	A_{l}=\bm{f}_{l}\bm{f}^{+}AD^{+},\]	
	and
\[A_{l}^{+}=DA^{+}\bm f \bm{f}_l^{+}.\]
	Due to the compatibility of 2-norm, there is
	\[\|A_l\|_{2}\le \|A\|_{2} \cdot \|\bm{f}^{+}\|_{2} \cdot \|\bm{f}_l\|_{2},\]
	and
	\[\|A_l^{+}\|_{2}\le \|A^{+}\|_{2} \cdot \|\bm f\|_{2} \cdot \|\bm{f}_l^{+}\|_{2}.\]
	Based on the above two inequalities, we have
	\[cond(A_{l})\le cond(A)cond(\bm f)cond(\bm{f}_{l}).\]
where $cond(\cdot)$ represents the generalized condition number. Since the Moore-Penrose generalized inverse matrix gives $\bm{f}^{+}= \frac{\bm{f}^T}{\|\bm f\|_{2}^2}$, it holds $ \|\bm f\|_{2} \cdot \|\bm f^{+}\|_{2} = 1 $.
Therefore, we have
\[cond(\bm{f}_l) = cond(\bm f)  = 1.\]
Finally, we arrive at
	\[cond(A_{l})\le cond(A),\]
	which completes the proof.
\end{IEEEproof}
Since large condition numbers lead to numerical instability and severe sensitivity to noisy measurements \cite{Jorgensen2013}, the low-resolution image can be used as prior to improve the solution of limited-angle CT reconstruction. Considering this, we propose the following constrained minimization problem for CT reconstruction
\begin{equation}\label{minimization problem}
	\begin{aligned}
		\min_{u} \quad &\mathcal F(A\bm u,f)+\mathcal R(\bm u)\\
		\mbox{s.t.} \quad &D\bm u=\bm u_l,
	\end{aligned}
\end{equation}
where $\mathcal F(\cdot)$ denotes the data fidelity term, and $\mathcal R(\cdot)$ denotes the regularization term.

\begin{figure*}[!thb]
	\centering
	\includegraphics[width=7.1in]{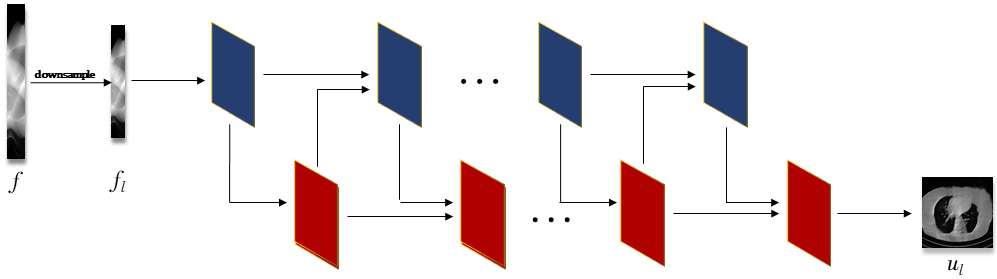}
	\caption{The network of low-resolution reconstruction model. The down-sampled projection data is reconstructed using the learned primal-dual algorithm to obtain the low-resolution image, where the proximal operators
    have been replaced with CNNs, and each block is composed of three convolution layers, PReLU and residual connection.}
	\label{1}
\end{figure*}

\subsection{Learned alternating direction algorithm }
The constrained minimization problem \eqref{minimization problem} can be further
reformulated into an unconstrained minimization problem using
the penalty method as follows
\begin{align}\label{equation1}
	\min_{\bm u} \quad &\mathcal F(A\bm u,\bm f)+\mathcal R(\bm u)+\frac{1}{2\mu}\left\|D\bm u-\bm u_l\right\|_2^2,
\end{align}
where $\mu$ is a positive regularization parameter. Because all terms in \eqref{equation1} contain the variable $\bm u$ making the minimization task difficult, we introduce
an auxiliary variable $\tilde{\bm{u}}$ and rewrite \eqref{equation1} as the following minimization problem
\begin{align*}
	\min_{\bm u,\tilde{\bm{u}}} \mathcal F(A\tilde{\bm{u}},\bm f)+\mathcal R(\tilde{\bm{u}})+\frac{1}{2\mu}\left\|D\bm u-\bm u_l\right\|_2^2+\frac{1}{2r}\left\|\tilde{\bm{u}}-\bm u\right\|_2^2,
\end{align*}
where $r$ is a positive parameter. The main advantage of above minimization problem is that we can use the alternating direction method to solve the multi-variable minimization problem. The variables $\tilde{\bm{u}}$ and $\bm u$ can be estimated alternatively by solving the following two energy minimization problems
\begin{align}\label{equation3}
	\min_{\tilde{\bm{u}}} \quad &\mathcal F(A\tilde{\bm{u}},\bm f)+\mathcal R(\tilde{\bm{u}})+\frac{1}{2r}\left\|\tilde{\bm{u}}-\bm u\right\|_2^2,
\end{align}
and
\begin{align}\label{equation4}
	\min_{\bm u} \quad &\frac{1}{2r}\left\|\tilde{\bm{u}}-\bm u\right\|_2^2+\frac{1}{2\mu}\left\|D\bm u-\bm u_l\right\|_2^2,
\end{align}
respectively.
The minimization \eqref{equation3} is a typical regularization model, which can be solved by the learned primal-dual algorithm as follows
\begin{equation*}\label{equation5}
	\left\{
	\begin{array}{l}
		\bm{p}^{k+1} = \mathop{\arg}\min\limits_{\bm p} \mathcal{F}^*(\bm p,\bm f)-\left\langle A\tilde{\bm{u}}^k,\bm p \right\rangle+\frac{1}{2\tau}\left\|\bm p-\bm{p}^k\right\|_2^2,\\
		\tilde{\bm{u}}^{k+1} = \mathop{\arg}\min\limits_{\tilde{\bm{u}}} \mathcal{R}(\tilde{\bm{u}})+\left\langle A\tilde{\bm{u}},\bm{p}^{k+1} \right\rangle+\frac{1}{2\tau}\left\|\tilde{\bm{u}}-\bm{u}^k\right\|_2^2,
	\end{array}
	\right.
\end{equation*}
where $\mathcal{F}^*$ denotes the adjoint operator of $\mathcal{F}$, $\bm p$ is the dual variable of $\tilde{\bm{u}}$, $\tau$ is a positive parameter. On the other hand, the minimization problem \eqref{equation4} is a least squared problem, which can be solved by the closed-form solution. To sum up, we propose to use the following iterative scheme to solve the minimization problem \eqref{equation3} and \eqref{equation4} in an alternative way
\begin{equation}\label{equation6}
	\left\{
	\begin{array}{l}
	    \bm u^{k+1}=(\mu\mathcal{I}+rD^*D)^{-1}(\mu\tilde{\bm u}^k,rD^*\bm u_l),\\
		\bm p^{k+1}=(\mathcal{I}+\tau\partial\mathcal{F}^*)^{-1}(\bm p^k, A\tilde{\bm u}^k,\bm f),\\
		\tilde{\bm u}^{k+1}=(\mathcal{I}+r\partial\mathcal{R})^{-1}(\bm u^k,A^*\bm p^{k+1}),\\
		
	\end{array}
	\right.
\end{equation}
where $\mathcal{I}$ denotes the identity operator, $A^*$ and $D^*$ represent the adjoint operators of the forward operator $A$ and down-sampling operator $D$, respectively. Now we intend to summarize the unrolled alternating direction algorithm by the deep neural networks to solve the low-resolution image prior based CT reconstruction model \eqref{minimization problem} as Algorithm 1, which is called the LRIP-net in the followings.
\vspace{0.2cm}
\renewcommand\arraystretch{1.5}
\hrule\hrule\hrule\hrule
\vspace{0.2cm}
{\noindent{\bf Algorithm 1} The Low-Resolution Image Prior based Network} \nolinebreak
\vspace{0.2cm}
\hrule\hrule
\vspace{0.1cm}
\begin{itemize}
	\item[1:] Initialize  $\bm{p}^0$, $\bm{u}^0,\tilde{\bm u}^0$
	\item[2:] \textbf{for} $k=0,$\dots$,I$, \textbf{do}
	\item[3:] \quad~ $\bm{u}^{k+1}\gets\Pi_{\theta^{\bm u}}(\tilde{\bm{u}}^{k},D^*\bm{u}_l)$;
	\item[4:] \quad~ $\bm p^{k+1}\gets\Gamma_{\theta^{\bm p}}(\bm p^k,A\tilde{\bm u}^k,\bm f)$;
	\item[5:] \quad~ $\tilde{\bm u}^{k+1}\gets\Lambda_{\theta^{\tilde{\bm u}}}(\bm{u}^{k+1},A^*\bm{p}^{k+1})$;
	\item[6:] \textbf{return} $\bm{u}^I$
\end{itemize}
\vspace{0.2cm}
\hrule\hrule\hrule\hrule
\vspace{0.15cm}
\begin{remark}
We assume the constraint $\tilde{\bm u}=\bm u$ holds unconditionally during the iterations process, where $\bm{u}^{k+1}$ is used to update $\tilde{\bm{u}}^{k+1}$ in the algorithm.
\end{remark}

\section{Algorithm implementation}
\label{sec3}
Our Low-Resolution Image Prior based Network (denoted by LRIP-net) is generated based on Algorithm 1, which is implemented in Python using Operator Discretization Library (ODL), the package Adler, the ASTRA Toolbox, and Tensorflow 1.8.0. Tensorflow is a toolkit for dealing with complex mathematical problems, it can be thought of as a programming system in which you represent calculations as graphs, mathematical operations as nodes, and communication multidimensional data arrays as edges of graphs. ASTRA
toolbox is a MATLAB and Python toolbox of high-performance
GPU primitives for 2D and 3D tomography, the ODL is a Python library for fast prototyping focusing on inverse problems and the Adler is a toolkit that can quickly implement neural network construction.
\begin{figure*}[!thb]
	\centering
	\includegraphics[width=7.1in]{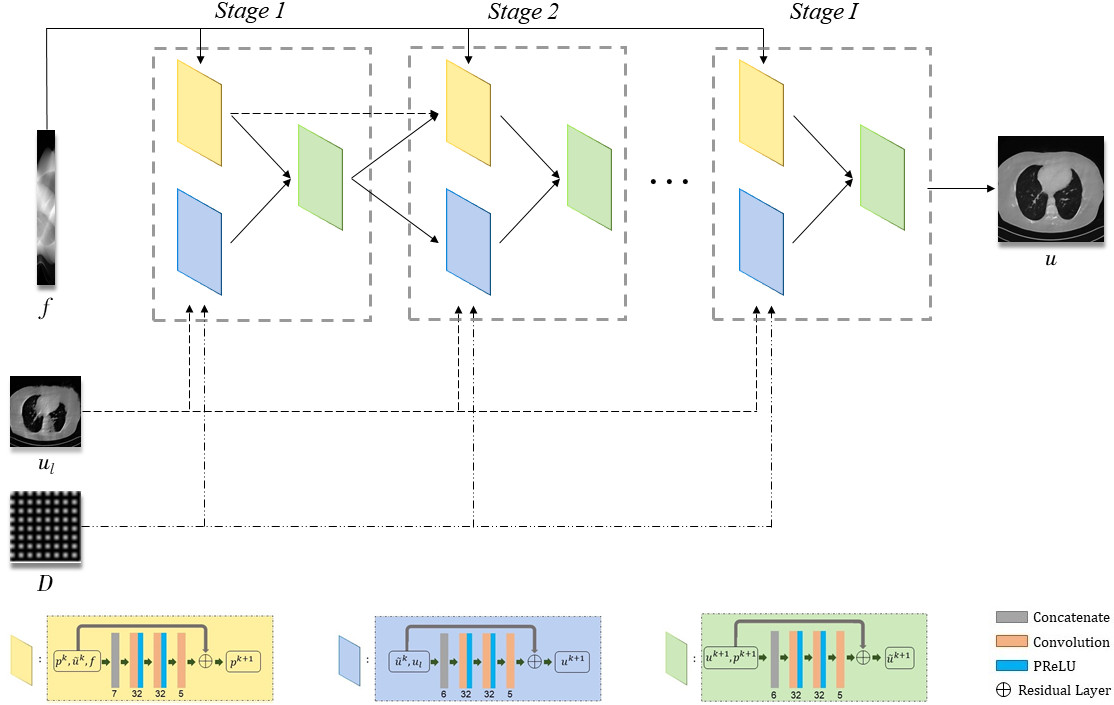}
	\caption{The network structure of our LRIP-Net: low-resolution image prior based reconstruction model. The reconstructed image is estimated by an iterative unrolling algorithm, and each block corresponds to a variable update. The concrete structure of each block is presented at the bottom.}
	\label{2}
\end{figure*}

\subsection{Network architecture}
The unrolling strategy is a discriminative learning method by unrolling an iterative optimization algorithm into a hierarchical architecture. Fig. \ref{1} depicts the network structures of low-resolution reconstruction model, we use the classical learned primal-dual reconstruction method \cite{adler2018learned} to obtain the low-resolution solution $\bm{u}_l$,  for which we can also adopt other advanced learning-based methods or variational methods. Fig. \ref{2} depicts the network structures of our LRIP-net, which has three inputs including the incomplete projection data $\bm f$, the system matrix $A$, and the reconstructed low-resolution image $\bm{u}_l$. More specifically, there are three blocks in each stage of the high-resolution reconstruction, which correspond to the three variables. As shown at the bottom of Fig. \ref{2}, each block involves a 3-layer network.

The total depth of the network depends on the number of stages contained in the network, which is chosen to balance the receptive fields and the total number of parameters. The proposed network introduces the residual structure for two reasons: 1) the residual structure makes the network easier to train and optimize because each update is only a small offset, and 2) the skip connections can alleviate gradient disappearance and gradient explosion caused by increasing the depth of deep neural networks. The non-linear activation functions are chosen as the Parametric Rectified Linear Units (PReLU) function.
As displayed in Fig. \ref{2}, we set the numbers of channels in each stage as $7\rightarrow32\rightarrow32\rightarrow5$ for $\bm p$, $6\rightarrow32\rightarrow32\rightarrow5$ for $\tilde{\bm u}$ and $\bm u$,
where the differences in the numbers are due to the dimension of inputs. The convolutions are set to be of the size $3 \times 3$ in our network. Furthermore, we choose the Xavier initialization scheme for the convolution parameters and the zero initialization for all biases. The convolution stride is set as 1 and the padding strategy is chosen as `SAME' in the network.

\begin{figure*}[th]
	\begin{center}
		\subfloat[Loss function]{\includegraphics[width=5.8cm,height=5.0cm]{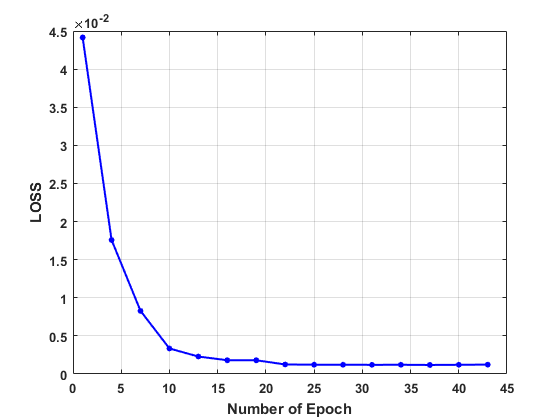}}\hspace{-2ex}
		\subfloat[PSNR]{\includegraphics[width=5.8cm,height=5.0cm]{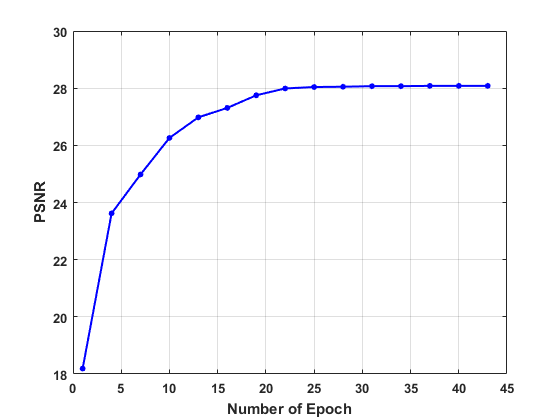}}\hspace{-2ex}
		\subfloat[SSIM]{\includegraphics[width=5.8cm,height=5.0cm]{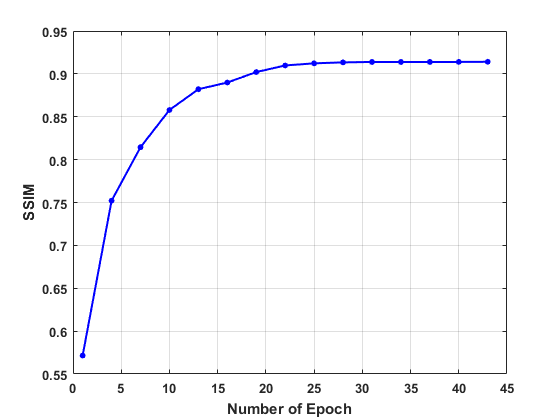}}\hspace{-2ex}
		\caption{The values of loss function, PSNR and SSIM with respect to the numbers of  epochs in our network, where the curves are evaluated on the human phantom with 90$^{\circ}$ scanning angular range and 5\% Gaussian noise.}
		\label{psnr_ssim}
	\end{center}
\end{figure*}

\subsection{Network loss and optimization}
In the training stage, we use both Mean Squared Error (MSE) and Structural Similarity Index (SSIM) as our loss function defined below
\begin{equation}\label{loss1}
	\begin{aligned}
		\mathcal{L} = \frac{1}{L}\sum\limits_{i=1}^L\Big(\mathcal L_{MSE} \big({\bm u_i}, {\bm u_i^*}\big) + \mu\mathcal{L}_{SSIM}\big(\bm u_i, \bm u_i^*\big)\Big),
	\end{aligned}
\end{equation}
where $\bm u_i$ denotes the reconstructed image, $\bm u_i^*$ denotes the reference image and $\mu$ is a trade-off parameter. We assumed that both the MSE loss and SSIM loss have the same contribution. Thus, $\mu$ is fixed as $\mu =1$ for all experiments.

Our network updates each parameter through the backpropagation of the stochastic gradient descent method in Tensorflow. For a fair comparison, most experimental parameters are set as the same as the PD-net and FSR-net. We adopt the adaptive moment estimation (Adam) to optimize the learning rate by setting the parameter $\beta = 0.99$ and other parameters to their default values. The learning rate schedule is set according to cosine annealing to improve training stability, the initial learning rate $\eta_0$ is set to $10^{-4}$. To further improve the training stability, the global gradient norm clipping is performed by limiting the gradient norm to 1. Besides, the batch size is set to 1 for all experiments.

\section{NUMERICAL RESULTS}
\label{sec4}
In this section, we evaluate our LRIP-net on limited-angle reconstruction problems and compare it with several state-of-the-art methods on a human phantom dataset.
\subsection{Comparison algorithms}
We adopt several recent CT reconstruction methods, both the variational method and learning-based methods as given below
\begin{itemize}
	\item TV model: \textit{the TV\ regularized reconstruction model} in \cite{2008Image}. We tuned the balance parameter $\lambda \in [0.9, 2.5]$, the step size for the primal value $\tau \in [0.5, 0.9]$ and the step size for the dual value $\sigma \in [0.2, 0.5]$ for different experiments.
	\item FBP-Unet: \textit{the FBP-Unet reconstruction} in \cite{Jin2017}. It is a method combining the FBP reconstruction with the Unet as the post-processing to improve image quality. We use the Xavier to initialize the network parameters. And the loss function is the mean squared loss of the reconstructed image and the ground truth.
	\item PD-net: \textit{the Learned Primal-Dual network} in \cite{adler2018learned}. The network is a deep unrolled neural network with 10 stages. The number of initialization channels for primal values and dual values is set to 5. The Xavier initialization and the mean squared loss of the reconstructed image and the ground truth are used in all experiments.
	\item SIPID: \textit{the deep learning framework for Sinogram Interpolation and Image Denoising in \cite{8363862}}. The SIPID network can achieve accurate reconstructions through alternatively training the sinogram interpolation network and the image denoising network. The Xavier initialization and the mean squared loss of the reconstructed image and the ground truth are used in all experiments.
	\item FSR-net: \textit{the Learned Full-Sampling Reconstruction From Incomplete Data} in \cite{cheng2020learned}. The network is an iterative expansion method that used the corresponding full-sampling projection system matrix as a prior information. To be specific, the IFSR-net guarantees the invertibility of the system matrix, while the SFSR-net guarantees numerical stability. The number of initialization channels for primal values and dual values is set as 6 and 7, respectively. And the loss function is the mean square error of the image domain and the Radon domain with the weight  $\alpha$ being 1.
\end{itemize}

\subsection{Datasets and settings}
We use the clinical data ``The 2016 NIH-AAPM-Mayo Clinic Low Dose CT Grand Challenge"\cite{mccollough2017low}, which contains 10 full-dose scans of the ACR CT accreditation phantom. We select 9 data as the training profile and leave 1 data for the evaluation, resulting in 2164 images of size 512$\times$512 for the training and 214 images for the testing. We concern with the limited-angle reconstruction problem in the numerical experiments, for which the scanning angular interval is set as 1 degree. The additive white Gaussian noises and Poisson noises are introduced into the projected data to validate the performance of reconstruction methods.

\subsection{Parameter behavior}
In the first place, we test the effect of the number of epoch on the convergence in network training on the human dataset. The values of the loss function, PSNR, and SSIM are plotted in Fig. \ref{psnr_ssim}, which evidence the convergence of our LRIP-net. Accordingly to plots, we fix the number of the epochs as $k=22$ during the training in the following experiments.

Secondly, the amount of the parameters reflects the complexity of the network. Table \ref{model_parameter} lists the sizes of the parameters for the learning-based models. Because each stage in our model involves three 3-layer networks, our model has a total of 90 convolution layers, which gives $3.6 \times 10^{5}$ parameters. For the fairness of comparison, we set the stage number for PD-net  and FSR-net to be 20 and 10, respectively.

\renewcommand\arraystretch{1.1}
\begin{table}[t]
	\centering
	\caption{Comparison of the parameters among the learning-based methods. }
	\label{model_parameter}
	\setlength{\tabcolsep}{3mm}{
		\begin{tabular}{p{4.5cm}p{3cm}}
			\toprule[1pt]
			$\mathrm{Method }$                       & $\mathrm{Number \; of \; parameters  }$                \\
			\hline
			FBP-Unet                       & $10^7$                                           \\
			SIPID                          & $2 \times 10^{7}$                       \\
			PD-net                         &  $2.4 \times 10^{5}$                  \\
			FSR-net                       &  $4.9 \times 10^{5}$                   \\
			LRIP-net                        &   $3.6 \times 10^{5}$                    \\
			\bottomrule[1pt]
	\end{tabular}}
\end{table}

\renewcommand\arraystretch{1.25}
\begin{table}[!t]
	\centering
	\caption{Performance of our LRIP-net with respect to different settings of parameters. }
	\setlength{\tabcolsep}{3mm}
	\begin{tabular}{p{2.45cm}p{1.3cm}p{1.3cm}p{1.3cm}}
		\toprule[1pt]
		$\mathrm{Settings}$ & $\mathrm{PSNR}$ &$\mathrm{RMSE}$ &$\mathrm{SSIM}$ \\\hline
		$N_{p}=4, N_{d}=4$ &24.7973 &0.0537&0.8783 \\
		$N_{p}=5, N_{d}=5$ &$\textbf{24.9391}$&$\textbf{0.0528}$ &$\textbf{0.8858}$ \\
		$N_{p}=6, N_{d}=6$ &24.8532 &0.0532&0.8823 \\
		$N_{p}=7, N_{d}=7$ &24.8349 &0.0534&0.8816\\
		\bottomrule[1pt]
	\end{tabular}
	\label{Nprimal}
\end{table}

Finally, expanding the variable space is a common network optimization technique, which allows the model to retain some memory for the variables making the training process more stable such as $\bm u=\left[\bm u^{(1)}, \bm u^{(2)},\dots, \bm u^{(N_{p})}\right]$ and $\bm p=\left[\bm p^{(1)}, \bm p^{(2)},\dots, \bm p^{(N_{d})}\right]$. In Table \ref{Nprimal}, we explore the influence of the choices of $N_p$ and $N_d$ to reconstruction accuracy on $90^{\circ}$ limited-angle data. As can be seen, the best accuracy is achieved with $N_p=5$ and $N_d=5$, which are fixed for all experiments.

\subsection{Experiments on data with Gaussian noises}
In this subsection, we evaluate the performance of our LRIP-net and other methods on limited-angle raw data corrupted different noise levels. We let the default value of $\tau$ to be $\tau=1/2$ to obtain the low-resolution image prior unless otherwise specified. And the LRIP-net trained by the MSE loss function is denoted by LRIP-net$_{\rm{MSE}}$.

\renewcommand\arraystretch{1.25}
\begin{table}[h]
	\caption{Comparison on limited-angle data corrupted by 5\% Gaussian noises in terms of PSNR, RMSE, SSIM and run time (ms).}
	\centering
	\begin{tabular}{p{0.6cm}|p{0.7cm}|p{1.65cm}|p{0.85cm}|p{0.8cm}|p{0.7cm}|p{0.5cm}}\hline\hline
		$\mathrm{Noise}$&$\mathrm{N_{view}}$&$\mathrm{Method}$&$\mathrm{PSNR}$&$\mathrm{RMSE}$&$\mathrm{SSIM}$&$\mathrm{Time}$\\\hline
		\multirow{33}*{$5\%$}&\multirow{11}*{$150^{\circ}$}&   FBP        & 13.5911     &0.1162 &  0.4854  &   \textbf{776} \\
		&&   TV         & 25.8815     &0.0631&  0.8091  &   56532  \\
		&&   FBP-Unet   & 23.8923     &0.0793 &  0.8703  &   1033   \\
		&&  SIPID      & 30.3275     &0.0321&  0.9276  &  1372 \\
		&&   PD-net     & 30.3766     &0.0313&  0.9301  & 1113     \\
		&&   IFSR-net    & 30.8763&0.0296   &  0.9303  & 1429 \\
		&&   SFSR-net    & 30.9411    &0.0279&  0.9324  & 1582 \\
		&&   LRIP-net$_{\mathrm{MSE}}$     & 31.3745 &0.0256    &0.9404   & 1245 \\
		&&   LRIP-net$_{1/2}$     & 31.5957 &   0.0247  &0.9426   &1264  \\
		&&   LRIP-net$_{1/4}$ &32.7354 &0.0231 &0.9499 &1240 \\
		&&   LRIP-net$_{1/8}$ &\textbf{32.8775} &\textbf{0.0218} &\textbf{0.9516} &1231 \\
		\cline{2-7}
		&\multirow{11}*{$120^{\circ}$}&    FBP        & 13.4418    &0.1652&0.4008     &  \textbf{602}   \\
		&&   TV         & 23.5852     &0.0802& 0.7891    &  60032 \\
		&&   FBP-Unet   & 21.1637     &0.0981&  0.8072  &   1023   \\\
		&&   {SIPID}      & {27.0428}     &{0.0416}&{0.9024}    & {1346}   \\
		&&   PD-net     & 27.1539&0.0402   &  0.9037  & 1132     \\
		&&   IFSR-net    & 28.0441 &0.0385  &  0.9079  & 1417 \\
		&&   SFSR-net    & 28.3263  & 0.0372&  0.9103  & 1551 \\
		&&   LRIP-net$_{\mathrm{MSE}}$       & 29.1888 &0.0331&  0.9354  & 1242  \\
		&&   LRIP-net$_{1/2}$      & 29.2763   & 0.0326  & 0.9361    & 1256 \\
		&&   {LRIP-net$_{1/4}$} & {30.0991} & {0.0313} & {0.9385} & {1239} \\
		&&   {LRIP-net$_{1/8}$} & {\textbf{30.8746}} & {\textbf{0.0286}} & {\textbf{0.9399}} &{1227} \\
		\cline{2-7}
		&\multirow{11}*{$90^{\circ}$}&    FBP        & 13.0314    &0.2260& 0.3881    &  \textbf{430}   \\
		&&   TV         & 19.9501     &0.0997& 0.6918    &  54917 \\
		&&  FBP-Unet   & 18.5181      &0.1035&  0.7481  &   1015   \\
		&&   {SIPID}      &{22.7492}      &{0.0626} &{0.8626}   &{1363}    \\
		&&   PD-net     & 22.6047     &0.0638&  0.8612  & 1099     \\
		&&   IFSR-net     & 23.9399  &0.0572 &  0.8744  & 1382 \\
		&&   SFSR-net    & 24.2494    &0.0566&  0.8761  & 1525 \\
		&&   LRIP-net$_{\mathrm{MSE}}$      &24.9391 &0.0528 & 0.8858   &1218  \\
		&&   LRIP-net$_{1/2}$    & 25.1555    &0.0516 & 0.8893   & 1247\\
		&&   {LRIP-net$_{1/4}$} &{25.9716} &{0.0484} &{0.9041} &{1205} \\
		&&   {LRIP-net$_{1/8}$} & {\textbf{26.3553}} & {\textbf{0.0451}} & {\textbf{0.9117}} & {1186} \\\hline\hline
	\end{tabular}
	\label{gaussian0.05table}
\end{table}

Table \ref{gaussian0.05table} displays the quantitative results of different methods on limited-angle data corrupted by 5\% Gaussian noises. The indexes used for evaluation are PSNR, RMSE, SSIM, and running time. We observe that the reconstruction qualities of all methods decrease as the scanning angle shrinks. It can be found out that LRIP-net$_{\mathrm{MSE}}$ has obvious numerical advantage compared to other comparison algorithms. After introducing the SSIM loss into the loss function, the advantage of LRIP-net$_{1/2}$ is further improved. Therefore, we suggest to use the joint loss function in implementation. Our LRIP-net$_{1/2}$ can provide better reconstruction accuracy with 0.6 dB, 0.9 dB, 0.9 dB higher PSNR than the SFSR-net for $150^{\circ}$, $120^{\circ}$, and $90^{\circ}$ reconstruction problem, respectively. What is even more important, compared to the SFSR-net, which is a dual-domain reconstruction method introducing the full-sampling system matrix as the prior knowledge, our model not only improves the reconstruction quality but also saves the computational time, which is also important for the practical application.

\begin{figure*}[t]
	\begin{center}
		{   \includegraphics[width=2.45cm,height=2.45cm]{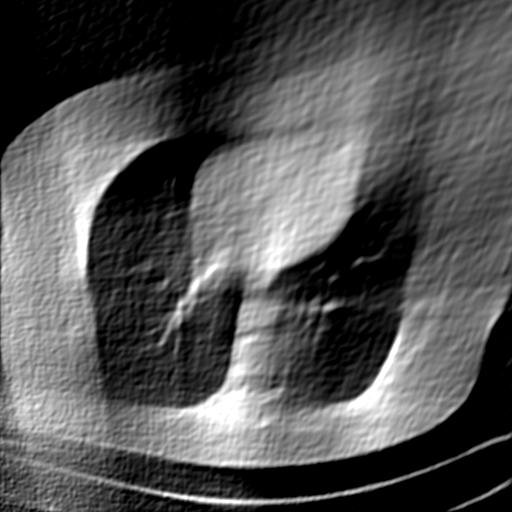}
			\includegraphics[width=2.45cm,height=2.45cm]{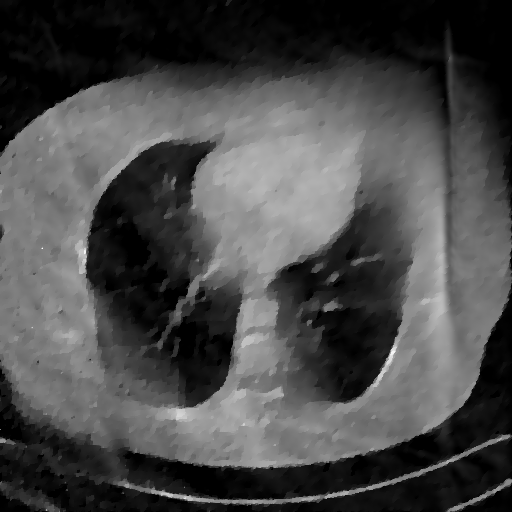}
			\includegraphics[width=2.45cm,height=2.45cm]{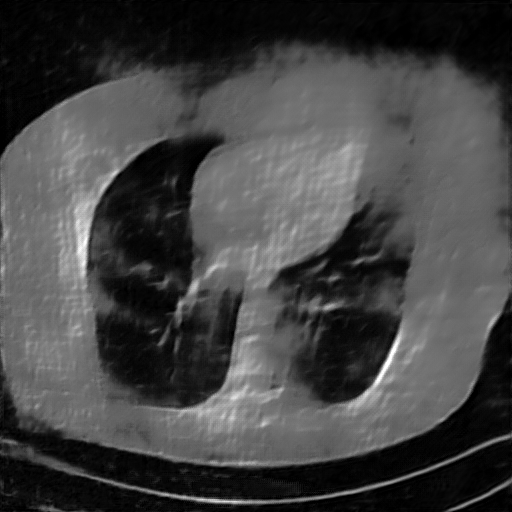}
			\includegraphics[width=2.45cm,height=2.45cm]{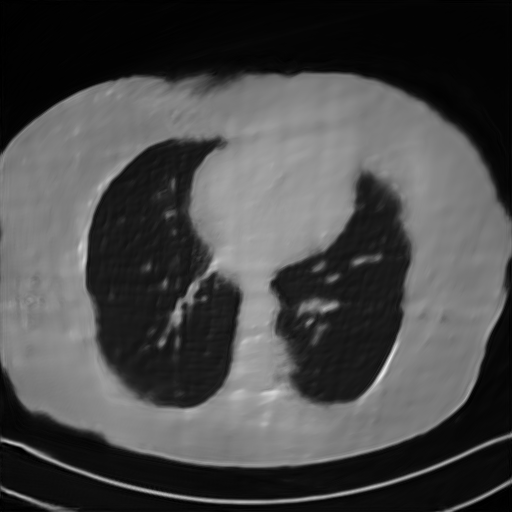}
			\includegraphics[width=2.45cm,height=2.45cm]{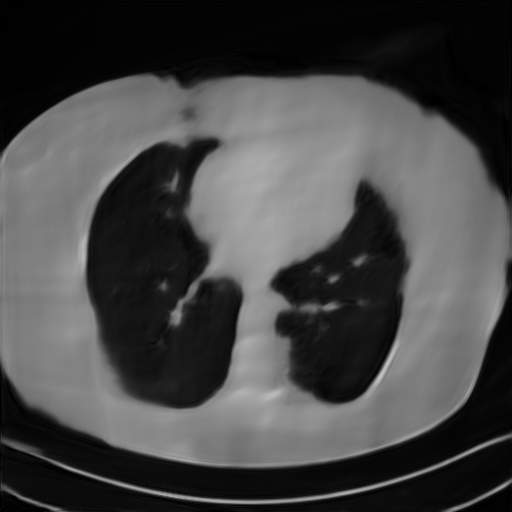}
			\includegraphics[width=2.45cm,height=2.45cm]{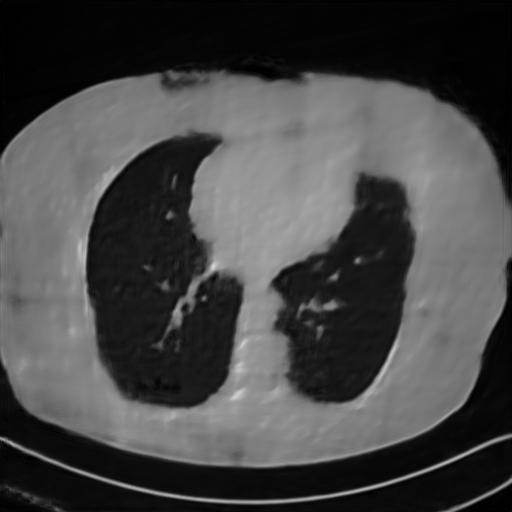}
			\includegraphics[width=2.45cm,height=2.45cm]{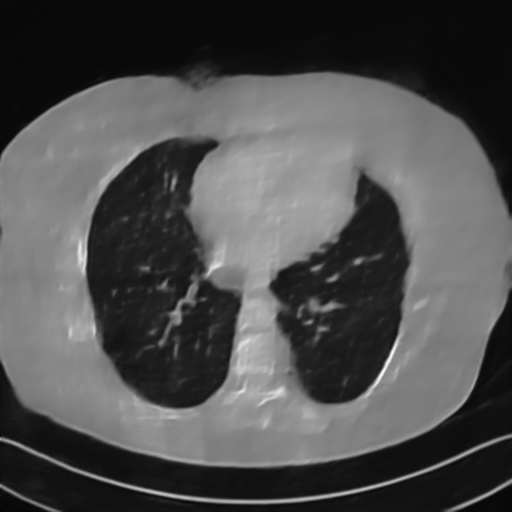}}\\
		\subfloat[FBP]{\includegraphics[width=2.45cm,height=2.45cm]{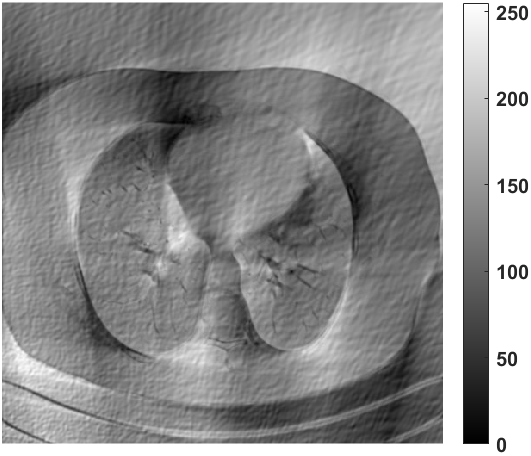}}
		\subfloat[TV]{\includegraphics[width=2.45cm,height=2.45cm]{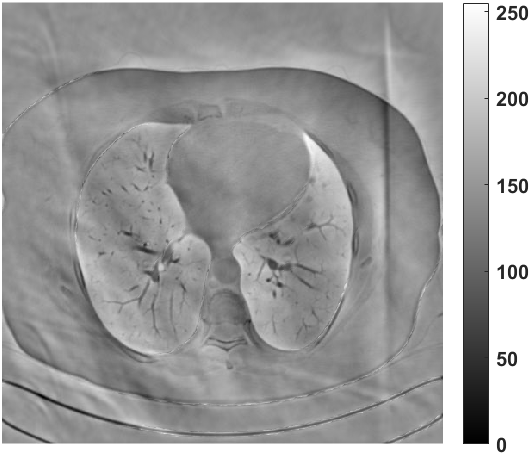}}
		\subfloat[FBP-Unet]{\includegraphics[width=2.45cm,height=2.45cm]{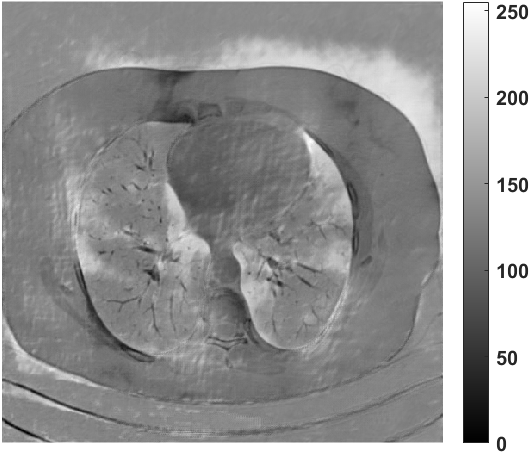}}
		\subfloat[SIPID]{\includegraphics[width=2.45cm,height=2.45cm]{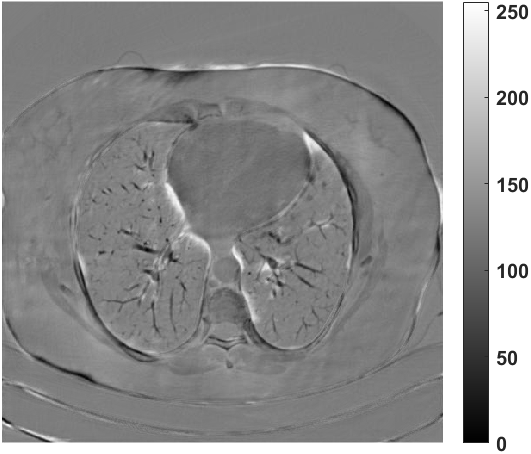}}
		\subfloat[PD-net]{\includegraphics[width=2.45cm,height=2.45cm]{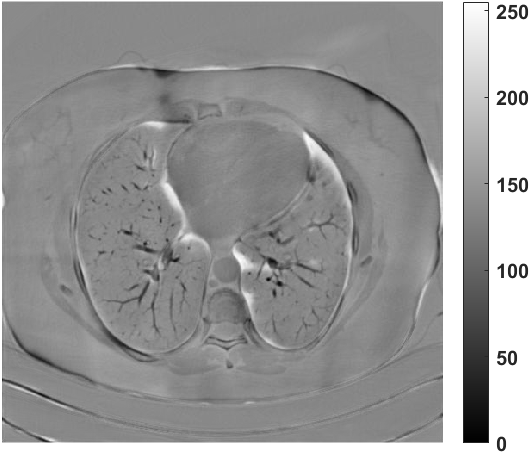}}
		\subfloat[SFSR-net]{\includegraphics[width=2.45cm,height=2.45cm]{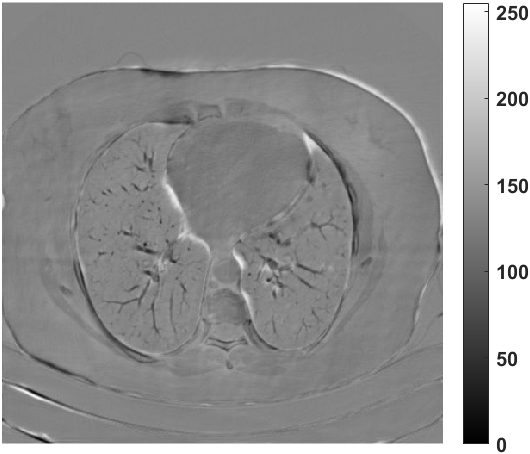}}
		\subfloat[LRIP-net$_{1/2}$]{\includegraphics[width=2.45cm,height=2.45cm]{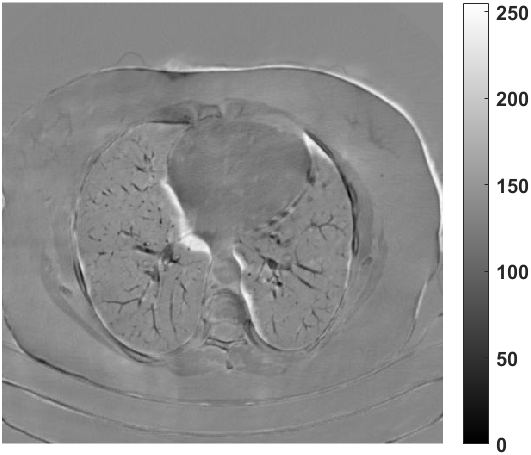}}
		\caption{\label{gaussian0.05fig}Limited-angle reconstruction experiment of the AAPM phantom dataset with 90$^{\circ}$ scanning angular range and 5\% Gaussian noises. Top row: the reconstructed images by different methods. Bottom row: the associated residual images. The display window is set as $\left[0,1\right]$.}
	\end{center}
\end{figure*}

Fig. \ref{gaussian0.05fig} presents the reconstruction results and residual images obtained by different methods for $90^{\circ}$ limited-angle reconstruction. As can be seen, the learning-based methods outperform the direct method and TV model, which exhibit serious artifacts in the missing angle region. Although the denoiser introduced by the FBP-Unet can somehow deal with the noises, the result still presents obvious artifacts. Compared to the SIPID, PD-net and FSR-nets, our LRIP-net$_{1/2}$ can better preserve the image details and edges with less information left in the residual images. Thus, both the quantitative and qualitative results confirm that the low-to-high double-resolution strategy can improve the reconstruction accuracy for the limited-angle reconstruction problem.

\begin{figure*}[!htb]
	\begin{center}
	    {\subfloat[LRIP-net$_{1/2}$]{
	    \includegraphics[width=2.85cm,height=2.85cm]{hl90ours005.png}
	    \includegraphics[width=2.9cm,height=2.9cm]{resDR12.png}}
	    \subfloat[LRIP-net$_{1/4}$]{\includegraphics[width=2.85cm,height=2.85cm]{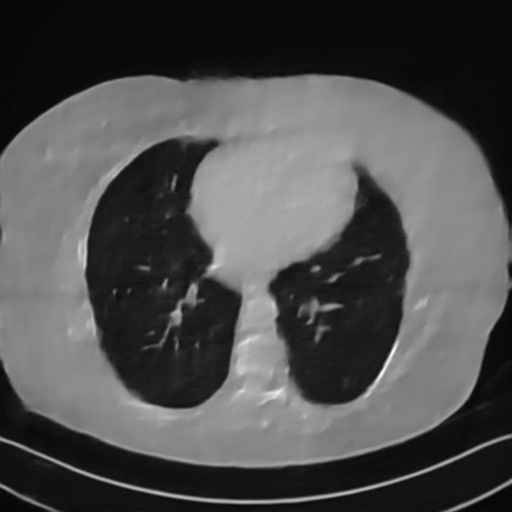}
	    \includegraphics[width=2.9cm,height=2.9cm]{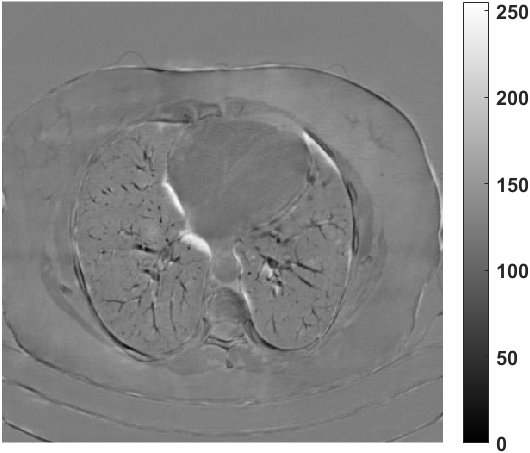}}
	    \subfloat[LRIP-net$_{1/8}$]{\includegraphics[width=2.85cm,height=2.85cm]{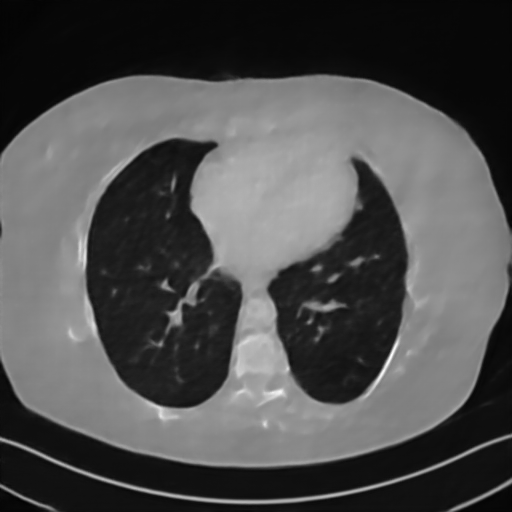}
	    \includegraphics[width=2.9cm,height=2.9cm]{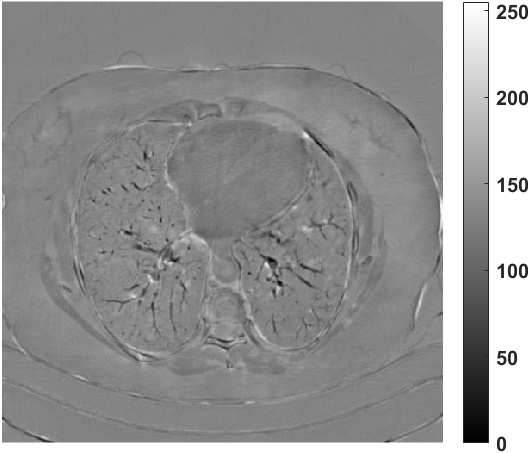}}
	    }
	    \caption{\label{gau1418fig}{Both the reconstruction images and residual images obtained by different image priors for limited-angle problem with 5\% Gaussian noises and 90$^\circ$ scanning angular range. The display window is set as $\left[0,1\right]$.}}
    \end{center}
\end{figure*}

We observe that the low-resolution image prior plays an important role in our method. More specifically, we compare the results of our LRIP-net with respect to different low-resolution priors, which are obtained by down-sampling rate of 1/2, 1/4, and 1/8, respectively. As can be seen in Table \ref{gaussian0.05table}, the best reconstruction results are obtained with the image prior reconstructed by the down-sampling rate of $1/8$ for 150$^\circ$, 120$^\circ$ and 90$^\circ$ limited-angle reconstruction. The visual comparison based on different image priors are also provided in Fig. \ref{gau1418fig}, where obviously less artifacts are left in the reconstruction image by LRIP-net$_{1/8}$.
By comparing the running time, it is easy to see that the smaller the low-resolution image prior, the faster the LRIP-net works.

\renewcommand\arraystretch{1.25}
\begin{table}[t]
	\caption{Comparison on limited-angle data corrupted by 10\% Gaussian noises in terms of PSNR, RMSE, SSIM and run time (ms).}
	\centering
	\begin{tabular}{p{0.6cm}|p{0.7cm}|p{1.65cm}|p{0.85cm}|p{0.8cm}|p{0.7cm}|p{0.5cm}}\hline\hline
		$\mathrm{Noise}$&$\mathrm{N_{view}}$&$\mathrm{Method}$&$\mathrm{PSNR}$&$\mathrm{RMSE}$&$\mathrm{SSIM}$&$\mathrm{Time}$\\\hline
		\multirow{33}*{$10\%$}&\multirow{11}*{$150^{\circ}$} &FBP&17.5464&0.1326&0.3914&\textbf{640}\\
		& &TV&20.7502&0.0917&0.5070&54536\\
		& &FBP-Unet&21.8293&0.0810&0.7887&1179\\
		& &{SIPID} &{29.0276}   &{0.0345}   &{0.9193}   &{1294}  \\
		& &PD-net&29.0084&0.0354&0.9193&1152\\
		& &SFSR-net&29.4543&0.0336&0.9199&1631\\
		& &IFSR-net&29.6694&0.0328&0.9231&1587\\
		& &LRIP-net$_{\mathrm{MSE}}$ &30.1257&0.0304&0.9283&1390\\
		& &LRIP-net$_{1/2}$&30.3367&0.0295 &0.9335 &1407 \\
		& &{LRIP-net$_{1/4}$}&{30.7131} &{0.0291} &{0.9359} &{1386} \\
		& &{LRIP-net$_{1/8}$}&{\textbf{30.8026}} &{\textbf{0.0288}} &{\textbf{0.9362}} &{1375} \\
		\cline{2-7}
		&\multirow{11}*{$120^{\circ}$}
		& FBP&14.8909&0.1801&0.2940&\textbf{558}\\
		& &TV&18.4345&0.1065&0.4365&55592\\
		& &FBP-Unet&20.0065&0.0999&0.7465&1207\\
		& &{SIPID} &{26.6271}   &{0.0461}   &{0.8941}   &{1311}  \\
		& &PD-net&26.7667&0.0458&0.8944&1221\\
		& &SFSR-net&27.2079&0.0436&0.9034&1624\\
		& &IFSR-net&27.2853&0.0432&0.9032&1539\\
		& &LRIP-net$_{\mathrm{MSE}}$&27.8333&0.0404&0.9083&1372\\
		& &LRIP-net$_{1/2}$  & 28.1961&0.0385  &0.9196 & 1389 \\
		& &{LRIP-net$_{1/4}$}&{28.4371} &{0.0369} &{0.9221} & {1366}\\
		& &{LRIP-net$_{1/8}$}&{\textbf{29.1261}} &{\textbf{0.0349}} &{\textbf{0.9256}} &{1361} \\
		\cline{2-7}
		&\multirow{11}*{$90^{\circ}$} &FBP&12.3727&0.2406&0.2455&\textbf{451}\\
		& &TV&15.9747&0.1956&0.4177&54633\\
		& &FBP-Unet&18.7582&0.1153&0.7252&1123\\
		& &{SIPID} &{23.6216}   &{0.0664}   &{0.8607}   &{1302}  \\
		& &PD-net&23.6473&0.0657&0.8615&1136\\
		& &SFSR-net&23.7253&0.0651&0.8591&1585\\
		& &IFSR-net&24.2056&0.0616&0.8701&1457\\
		& &LRIP-net$_{\mathrm{MSE}}$&24.5125&0.0597&0.8840&1339\\
		& &LRIP-net$_{1/2}$& 24.8153& 0.0576& 0.8975 & 1356\\
		& &{LRIP-net$_{1/4}$}&{24.9712} &{0.0562} &{0.8984} &{1327} \\
		& &{LRIP-net$_{1/8}$}&{\textbf{25.9377}} &{\textbf{0.0457}} &{\textbf{0.9141}} &{1319} \\
		\hline\hline
	\end{tabular}
	\label{gaussian0.1table}
\end{table}

\renewcommand\arraystretch{1.25}
\begin{table}[!ht]
	\caption{Comparison on limited-angle data corrupted by Poisson noises in terms of PSNR, RMSE, SSIM and run time (ms).}
	\centering
	\begin{tabular}{p{0.8cm}|p{1.7cm}|p{1.0cm}|p{0.9cm}|p{0.9cm}|p{0.8cm}}\hline\hline
		$\mathrm{N_{view}}$&$\mathrm{Method}$&$\mathrm{PSNR}$&$\mathrm{RMSE}$&$\mathrm{SSIM}$&$\mathrm{Time}$\\\hline
		\multirow{11}*{$150^{\circ}$}&FBP&19.1018&0.1108&0.6626&\textbf{692}\\
		&TV&28.4326&0.0378&0.8925&64075\\
		&FBP-Unet&23.2040&0.0691&0.8743&1134\\
		&{SIPID} &{28.7649}   &{0.0371}   &{0.9152}   &{1298}  \\
		&PD-net&28.8137&0.0362&0.9160&1288\\
		&IFSR-net&29.4316&0.0337&0.9236&1512\\
		&SFSR-net&30.0356&0.0314&0.9280&1649\\
		&LRIP-net$_{\mathrm{MSE}}$ &30.2227&0.0307&0.9349&1324\\
		&LRIP-net$_{1/2}$ &30.3342&0.0303  &0.9338  &1341\\
		&{LRIP-net$_{1/4}$}&{30.8961} &{0.0285} &{0.9366} &{1315} \\
		&{LRIP-net$_{1/4}$}&{\textbf{31.1291}} &{\textbf{0.0277}} &{\textbf{0.9368}} &{1308} \\
		\cline{1-6}
		\multirow{11}*{$120^{\circ}$}&FBP&15.8807&0.1606&0.5243&\textbf{599}\\
		&TV&25.3223&0.0541&0.8386&62783\\
		&FBP-Unet&19.5694&0.1050&0.8201&1119\\
		&{SIPID} &{27.4079}   &{0.0424}  &{0.9012}   &{1305}  \\
		&PD-net&27.4054&0.0426&0.9007&1220\\
		&IFSR-net&27.6385&0.0415&0.9107&1687\\
		&SFSR-net&27.5545&0.0419&0.9115&1752\\
		&LRIP-net$_{\mathrm{MSE}}$&27.7244&0.0411&0.9142&1351\\
		&LRIP-net$_{1/2}$   &27.7481&0.0409   &0.9157    &1367\\
		&{LRIP-net$_{1/4}$}&{28.4571} &{0.0374} &{0.9206} &{1327} \\
		&{LRIP-net$_{1/8}$}&{\textbf{29.5724}} &{\textbf{0.0332}} &{\textbf{0.9264}} &{1311} \\
		\cline{1-6}
		\multirow{11}*{$90^{\circ}$}&FBP&13.0830&0.2217&0.4747&\textbf{495}\\
		&TV&21.5935&0.1013&0.7908&63049\\
		&FBP-Unet&19.3228&0.1381&0.7854&1099\\
		&{SIPID} &{23.2017}   &{0.0693}  &{0.8552}   &{1304}  \\
		&PD-net&23.1127&0.0698&0.8546&1130\\
		&IFSR-net&23.5283&0.0666&0.8711&1560\\
		&SFSR-net&24.0277&0.0627&0.8733&1679\\
		&LRIP-net$_{\mathrm{MSE}}$&24.6883&0.0589&0.8803&1285\\
		&LRIP-net$_{1/2}$   &24.7223&0.0582  &0.8816     &1306\\
		&{LRIP-net$_{1/4}$}&{25.2347} &{0.0541} &{0.8892} &{1277} \\
		&{LRIP-net$_{1/8}$}&{\textbf{25.9344}} &{\textbf{0.0505}} &{\textbf{0.8981}} &{1265} \\
		\hline\hline
	\end{tabular}
	\label{possiontable}
\end{table}

\begin{figure*}[!htb]
	\begin{center}
		\subfloat[Clean image]{\includegraphics[width=4cm,height=4cm]{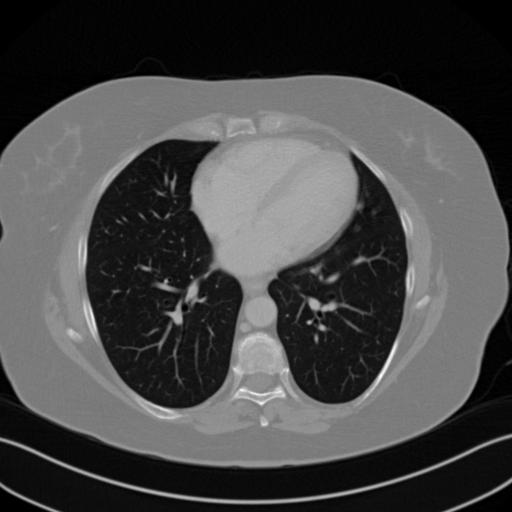}}
		\subfloat[FBP]{\includegraphics[width=4cm,height=4cm]{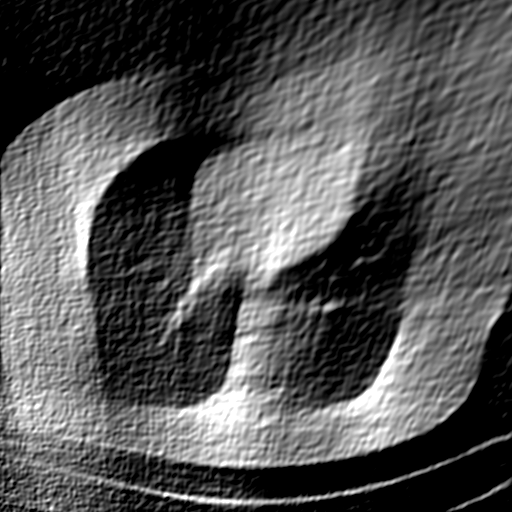}}
		\subfloat[TV]{\includegraphics[width=4cm,height=4cm]{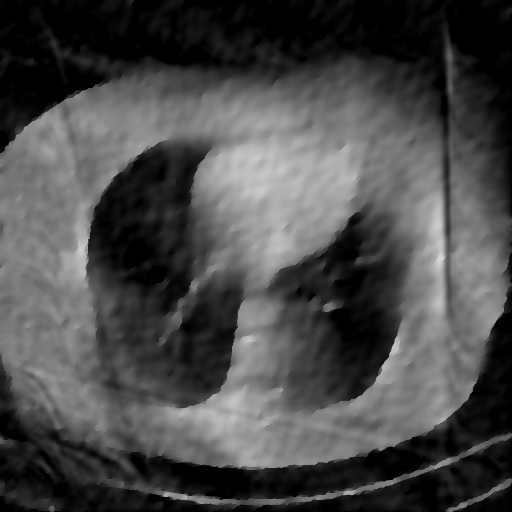}}
		\subfloat[FBP-Unet]{\includegraphics[width=4cm,height=4cm]{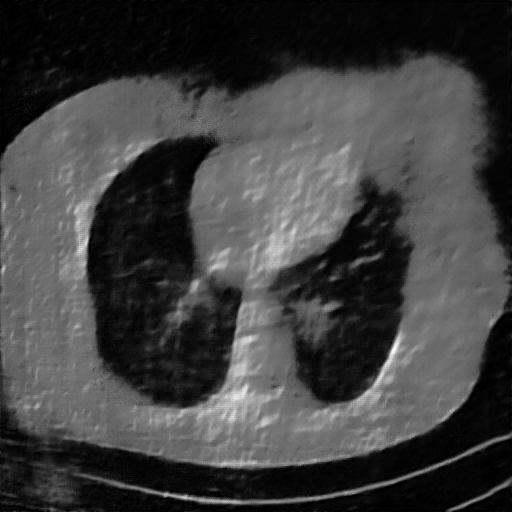}}
		\subfloat[SIPID]{\includegraphics[width=4cm,height=4cm]{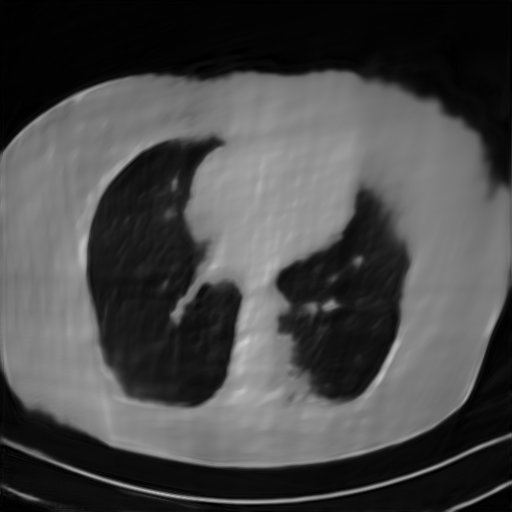}}
		\subfloat[PD-net]{\includegraphics[width=4cm,height=4cm]{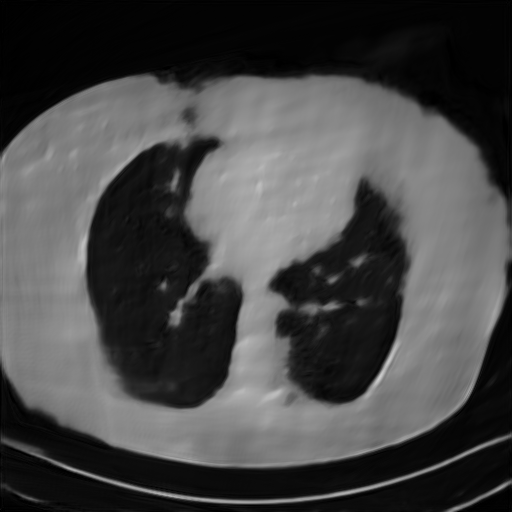}}
		\subfloat[IFSR-net]{\includegraphics[width=4cm,height=4cm]{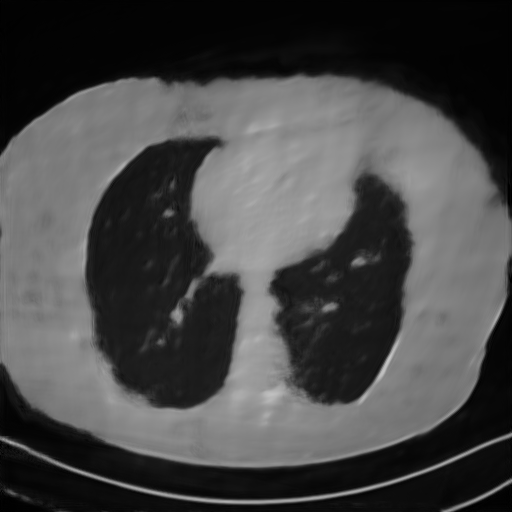}}
		\subfloat[SFSR-net]{\includegraphics[width=4cm,height=4cm]{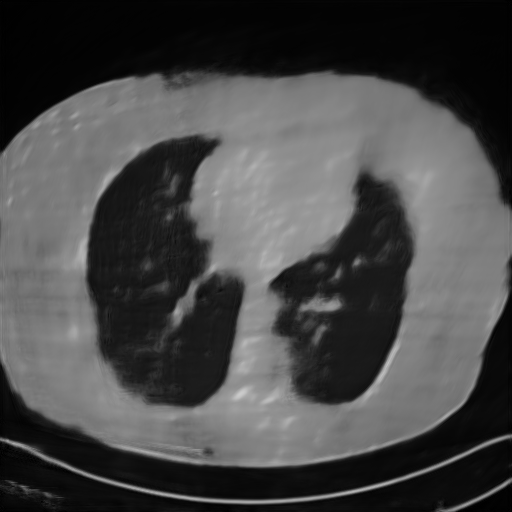}}
		\subfloat[LRIP-net$_{\mathrm{MSE}}$]{\includegraphics[width=4cm,height=4cm]{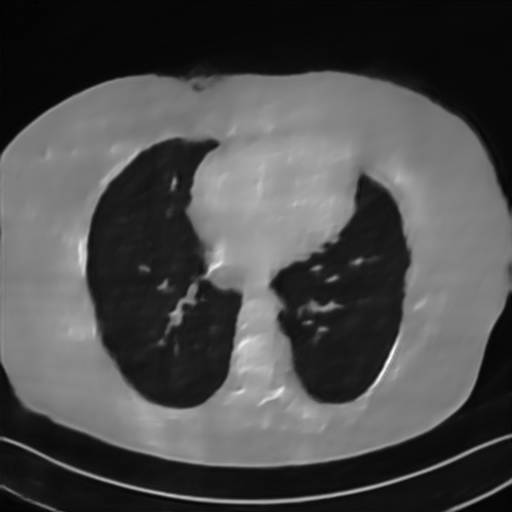}}
		\subfloat[LRIP-net$_{1/2}$]{\includegraphics[width=4cm,height=4cm]{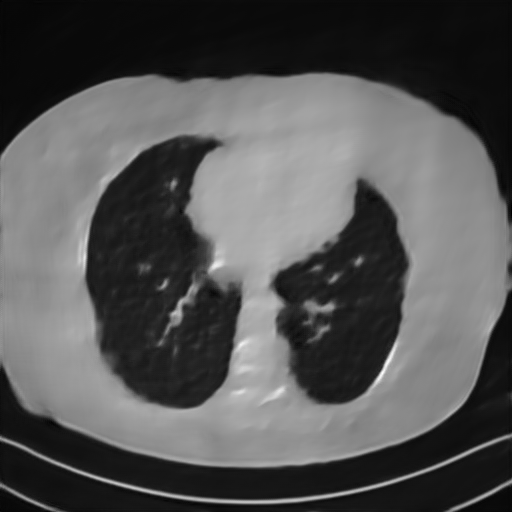}}
		\subfloat[LRIP-net$_{1/4}$]{\includegraphics[width=4cm,height=4cm]{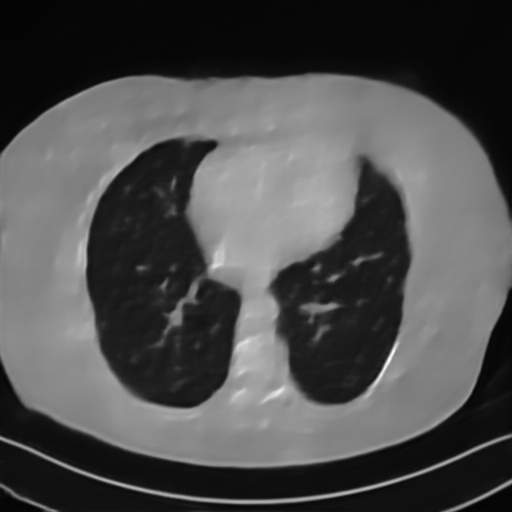}}
		\subfloat[LRIP-net$_{1/8}$]{\includegraphics[width=4cm,height=4cm]{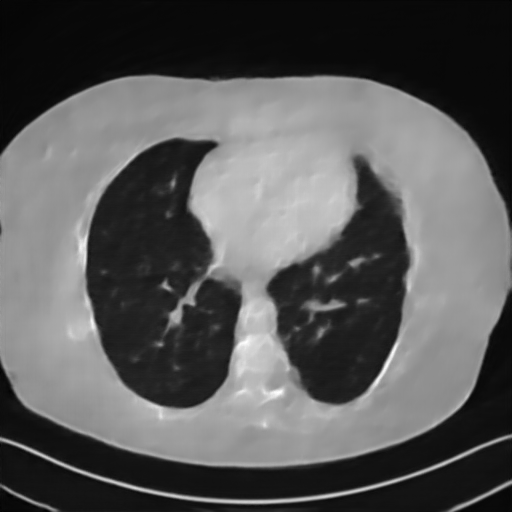}}
		\caption{Limited-angle reconstruction experiments on the AAPM phantom dataset within 90$^{\circ}$ scanning angular range and 10\% Gaussian noises. The display window is set as $\left[0,1\right]$. }
		\label{gaussian0.1fig}
	\end{center}
\end{figure*}

\begin{figure*}[!ht]
	\begin{center}
		\subfloat[Clean image]{\includegraphics[width=4cm,height=4cm]{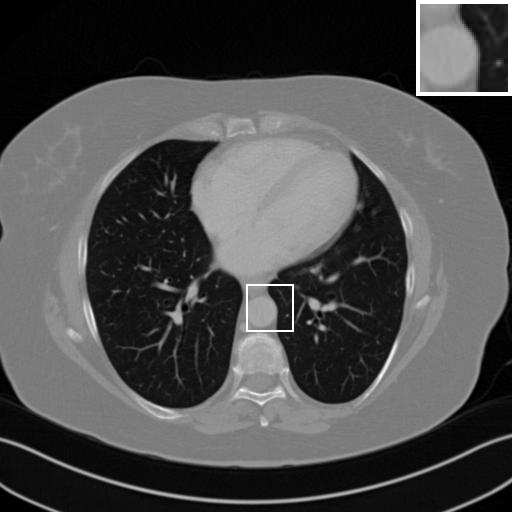}}
		\subfloat[FBP]{\includegraphics[width=4cm,height=4cm]{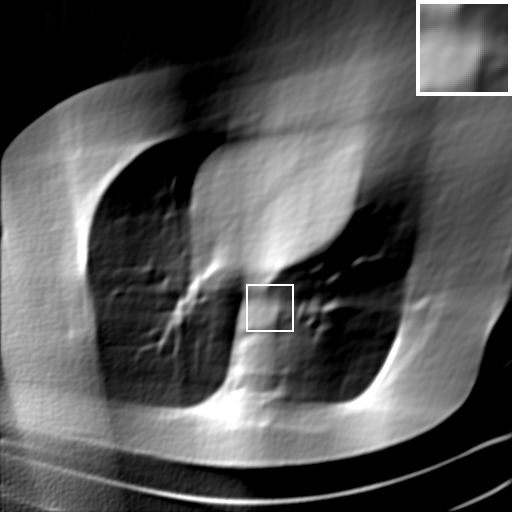}}
		\subfloat[TV]{\includegraphics[width=4cm,height=4cm]{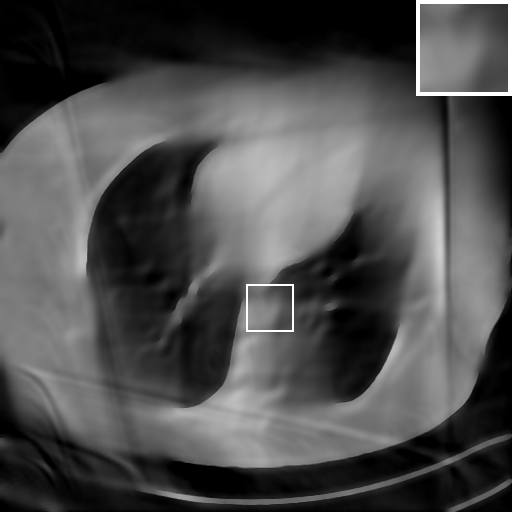}}
		\subfloat[FBP-Unet]{\includegraphics[width=4cm,height=4cm]{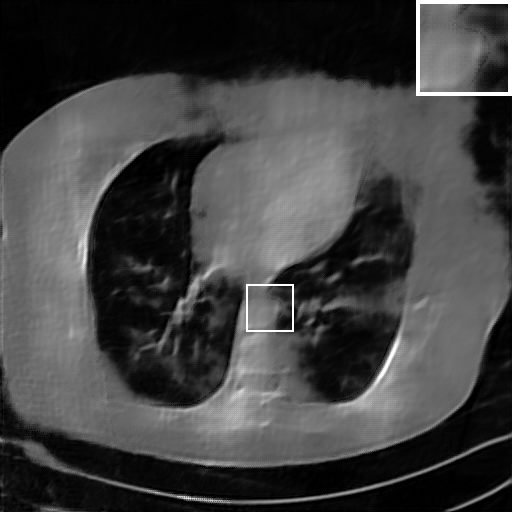}}
		\subfloat[SIPID]{\includegraphics[width=4cm,height=4cm]{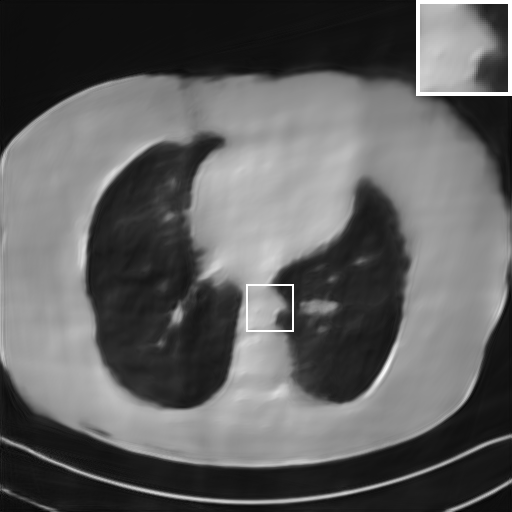}}
		\subfloat[PD-net]{\includegraphics[width=4cm,height=4cm]{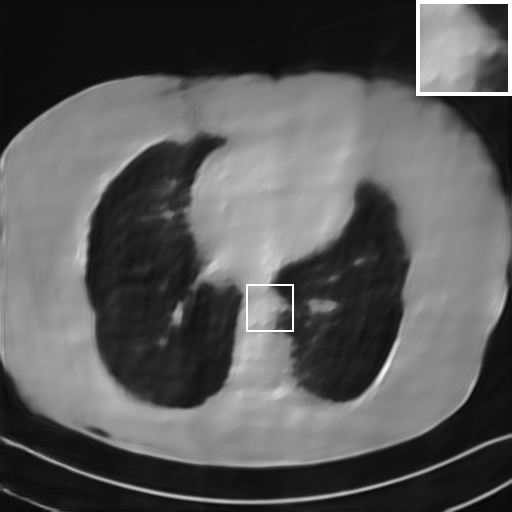}}
		\subfloat[IFSR-net]{\includegraphics[width=4cm,height=4cm]{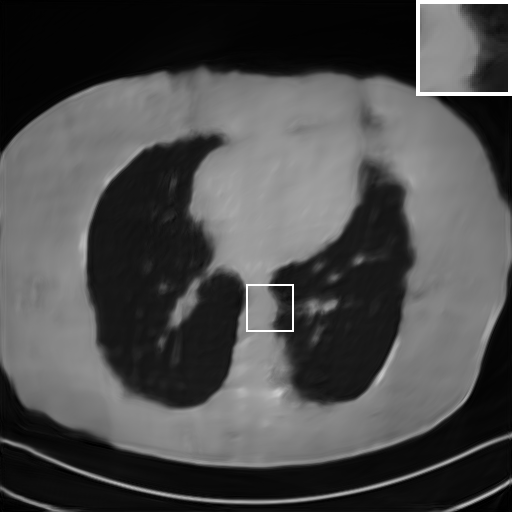}}
		\subfloat[SFSR-net]{\includegraphics[width=4cm,height=4cm]{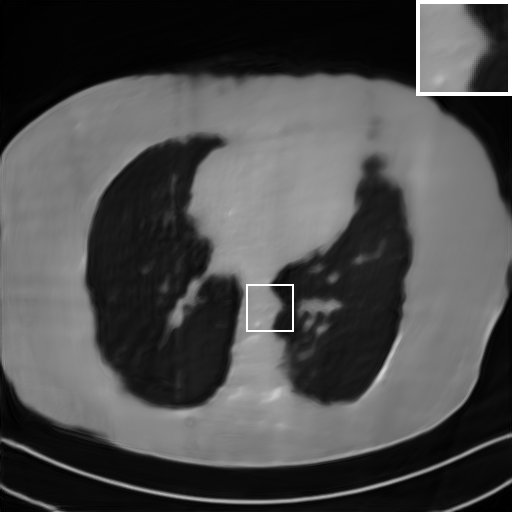}}
		\subfloat[LRIP-net$_{\mathrm{MSE}}$]{\includegraphics[width=4cm,height=4cm]{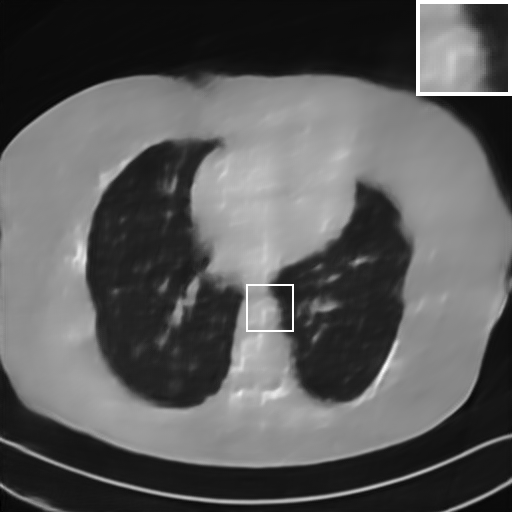}}
		\subfloat[LRIP-net$_{1/2}$]{\includegraphics[width=4cm,height=4cm]{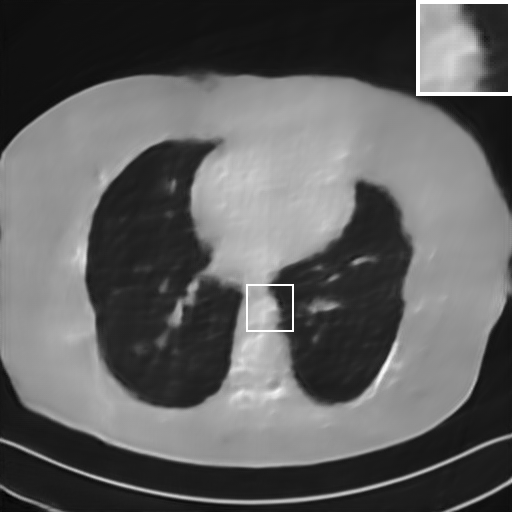}}
		\subfloat[LRIP-net$_{1/4}$]{\includegraphics[width=4cm,height=4cm]{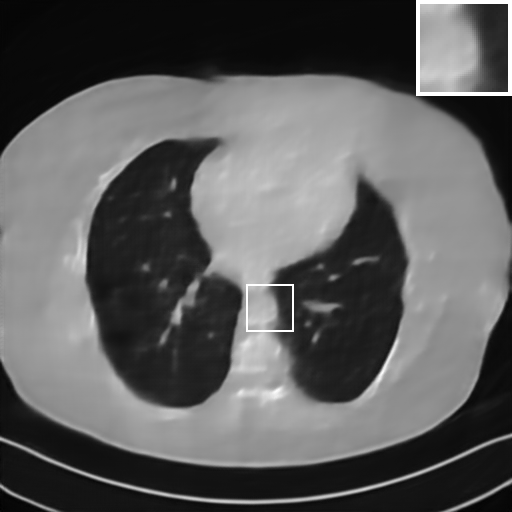}}
		\subfloat[LRIP-net$_{1/8}$]{\includegraphics[width=4cm,height=4cm]{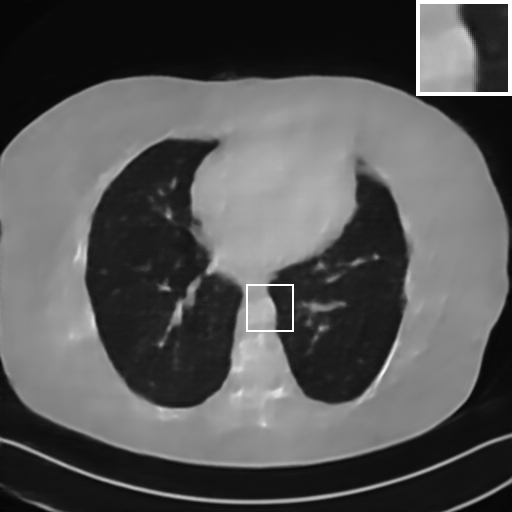}}
		\caption{Limited-angle reconstruction experiment on the AAPM phantom dataset with 90$^{\circ}$ scanning angular range and Poisson noises. The display window is set as  $\left[0,1\right]$.}
		\label{possionfig}
	\end{center}
\end{figure*}

We further increase the noise level contained in the raw data to 10\% white Gaussian noises and list the quantitative results in Table \ref{gaussian0.1table}. It can be observed that the reconstruction performance of the TV model is poor in the case of high-level noises with PSNR dropping by 4 to 5 dB compared to the previous experiments. On the other hand, the performance of the learning-based methods is less sensitive to noises. The SIPID method relying on the sinogram interpolation works better than FBP-Unet. And the deep unrolling methods (i.e., PD-net, IFSR-net, SFSR-net) outperform the traditional iterative algorithm when the scanning range is limited and data is corrupted by noises. Similar to the previous experiments, compared with other deep learning algorithms, our LRIP-nets give the reconstruction results with higher PSNR and SSIM. Moreover, the low-resolution image obtained by the projection data down-sampled with rate 1/8 always gives the best reconstruction results with more than 2 dB PSNR and 0.05 SSIM increments compared to the PD-net.
Fig. \ref{gaussian0.1fig} illustrates the reconstructed images from different methodologies with scanning angular range of $90^{\circ}$ and 10\% Gaussian noises.
It can be seen that the both TV model and the FBP-Unet suffers from significant artifacts, which present distortions in the angular range of the missing scan. Other learning-based methods provides better visual qualities than FBP-Unet, and our LRIP-net$_{1/8}$ still gives the best reconstruction result with correct boundaries and fine structures.

\renewcommand\arraystretch{1.25}
\begin{table*}[h]
	\caption{Comparison on limited-angle data corrupted by 5\% Gaussian noises in terms of PSNR, RMSE, and SSIM. }
	\centering
	\begin{tabular}{l|c|c|c|c|c|c|c|c|c}
	\hline
	\multirow{2}{*}{\diagbox[width=3cm]{$\mathrm{Method}$}{$\mathrm{Settings}$}}&\multicolumn{3}{c|}{150$^\circ$}&\multicolumn{3}{c|}{120$^\circ$}&\multicolumn{3}{c}{90$^\circ$}\\
		\cline{2-10}
		& PSNR&RMSE&SSIM& PSNR&RMSE&SSIM& PSNR&RMSE&SSIM \\
		\hline
		LRIP-net$_{1/2}$-FBP& 31.1571 &0.0267 &0.9374 &28.6398 &0.0369 &0.9279 &24.5943 &0.0545 &0.8814\\
		\hline
		LRIP-net$_{1/2}$-PDnet& 31.5957 &0.0247 &0.9426 &29.2763 &0.0326 &0.9385 &25.1555 &0.0516 &0.8893\\
	\hline
	
	\end{tabular}
	\label{table6}
\end{table*}

\subsection{Experiments on data with Poisson noises}
Due to the statistical error of low photon counts, Poisson noises are introduced and result in random
thin bright and dark streaks that appear preferentially along the direction of the greatest
attenuation \cite{Elbakri2002}. Table \ref{possiontable} lists the PSNR, RMSE, and SSIM of different methods on raw data scanned within a limited scanning angle and corrupted by Poisson noises, where Poisson noises correspond to 100 incident photons per pixel before attenuation. Unlike white Gaussian noises, the performance of the TV model is significantly better than the post-processing learning method FBP-Unet. And all other learning-based methods work better than the TV model in terms of PSNR, RMSE and SSIM. Our LRIP-nets still provide the best reconstruction accuracy among all the learning-based methods for the scanning angle of $150^{\circ}$, $120^{\circ}$, and $90^{\circ}$, respectively. It demonstrates that the LRIP-nets are also effective for data contaminated by Poisson noises.

Fig. \ref{possionfig} manifests the reconstruction results of these methods with scanning angular of $90^{\circ}$. It can be seen that both FBP and FBP-Unet produce serious artifacts within the range of missing angles. The TV model performs well in removing Poisson noises, but it can not handle the artifacts very well. Similarly, there left obvious artifacts on boundaries and different degrees of missing in visceral tissues of the reconstruction images obtained by the SIPID, PD-net and FSR-net. The visceral tissue and boundaries of our LRIP-net reconstructions are more intact and smoother, especially for the LRIP-net$_{1/8}$ which gives the ideal boundaries. The observation becomes even apparent if we look at the zoom-in regions, where the LRIP-nets can produce results with fine structures. Therefore, we conclude that the low-resolution image prior can effectively improve the qualities of the limited-angle CT reconstruction.

As far as the running time is concerned, since the FBP is an analytical reconstruction algorithm, it gives the fastest speed. On the other, the TV model is a traditional iterative method, for which the running time is the longest. For the deep learning-based methods, the running time increases with the complexity of the network, but the overall difference is not significant.

\begin{figure*}[h]
	\begin{center}
		\subfloat[Condition number by  1-norm]{\includegraphics[scale=0.4]{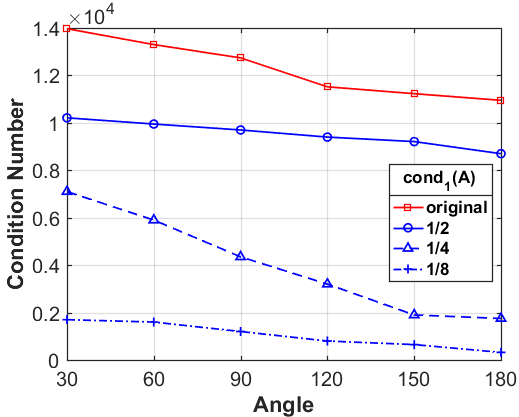}}\quad
		\subfloat[Condition number by  2-norm]{\includegraphics[scale=0.4]{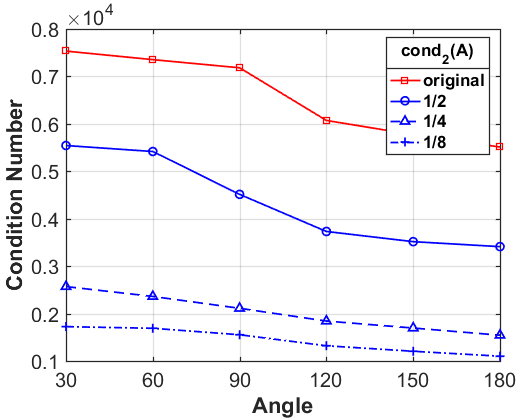}}\quad
		\subfloat[Condition number by $\infty$-norm]{\includegraphics[scale=0.4]{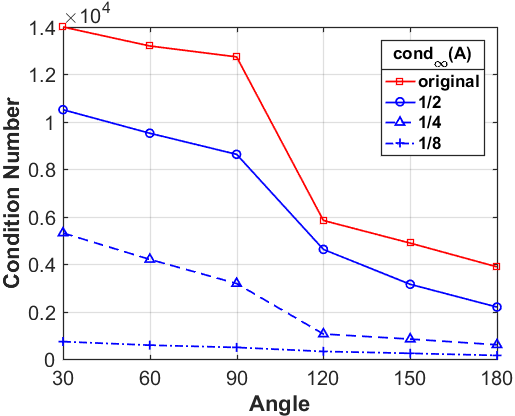}}
		\caption{The condition number of the system matrix corresponding to different sampling rat, where the down-sampling rate is used as 1/2, 1/4, and 1/8, respectively, and the number of bins is fixed as $N_{\rm{bins}}=736$.}
		\label{cond_fig}
	\end{center}
\end{figure*}

\begin{figure*}[t]
	\begin{center}
		\subfloat[PSNR]{\includegraphics[scale=0.4]{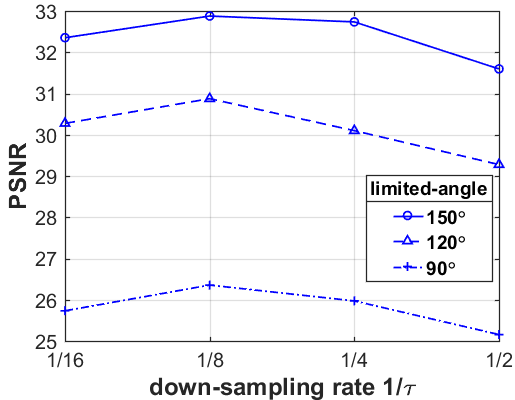}}\quad
		\subfloat[RMSE]{\includegraphics[scale=0.4]{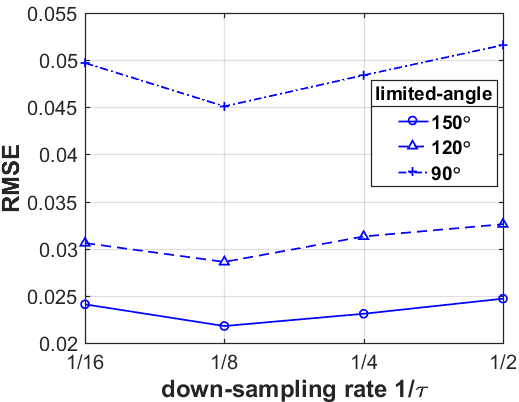}}\quad
		\subfloat[SSIM]{\includegraphics[scale=0.4]{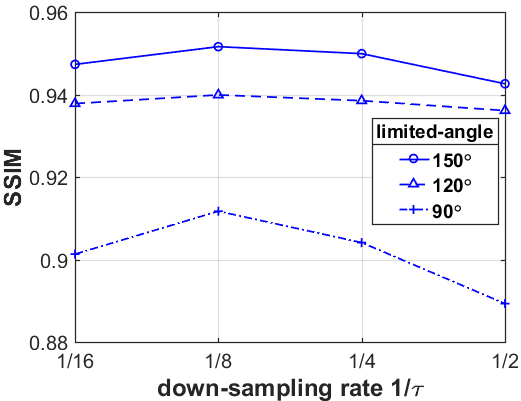}}
		\caption{{Comparisons of PSNR, RMSE and SSIM of the limited-angle reconstruction experiments on the AAPM phantom dataset corrupted by 5\% Gaussian noises, where the down-sampling rate is chosen as 1/2, 1/4, 1/8, and 1/16, respectively, to generate the low-resolution image prior.}}
		\label{fig9}
	\end{center}
\end{figure*}

\subsection{Discussions}

The primal-dual network can be regarded as our model without the low-resolution image prior. By comparing the results in Table \ref{gaussian0.05table}, \ref{gaussian0.1table} and \ref{possiontable}, even though both LRIP-net and PD-net are trained by the MSE loss, the LRIP-net shows significant advantages, always producing higher PSNR and SSIM. Through such ablation study, it is shown that the low-resolution image working as prior knowledge can effectively improve the qualities of the reconstruction images. On the other hand, we discuss the impact of the low-resolution image to our LRIP-net, where the high quality priors obtained by PD-net and poor quality priors obtained by FBP are used and compared in our LRIP-net.  As shown in Table \ref{table6}, two observations can be concluded as follows
\begin{itemize}
    \item Our LRIP-net with the FBP reconstructed image priors always produce better reconstruction results than PD-net (see Table \ref{gaussian0.05table}). Thus, the performance of our LRIP-net is not subject to the low-resolution image reconstruction method.
    \item Our LRIP-net with the PD-net reconstructed image priors always provide better reconstruction results than the LRIP-net using FBP reconstructed image priors. It means that the better the quality of a priori image, the better the reconstruction result.
\end{itemize}
Next, we numerically compare the condition numbers of the system matrices for both the low-resolution and original reconstruction problems by 1-norm, 2-norm and $\infty$-norm, respectively. As shown in Fig. \ref{cond_fig}, we can observe that the condition numbers of the low-resolution reconstruction problem are much smaller than the original reconstruction problems for different scanning angles, which are consistent with our theoretical analysis. In fact, larger condition numbers allow for undesired solutions, which also numerically satisfy the forward model \eqref{inverse problem}. More importantly, the constraint in our model \eqref{minimization problem} can help to find better solution from the null space, which has already been perfectly demonstrated by our numerical experiments.

What is more, as shown in Table \ref{gaussian0.05table}, \ref{gaussian0.1table} and \ref{possiontable}, the smaller the down-sampling rate, the better the reconstruction result. Since the quality of the image prior decreases as the resolution decreases for the limited-angle reconstruction problems, there is a trade-off between the resolution and the quality.
Thus, we track the relationship between the resolution of the low-resolution image priors and the performance of our LRIP-net. As shown in Fig. \ref{fig9}, the best reconstruction results are obtained using the low-resolution image prior with down-sampling rate $\tau=1/8$. Although the smaller the condition number of the system matrix, the better the numerical stability of the solution, it does not mean that the better the performance of our LRIP-net.

\section{Conclusions and Future works}
\label{sec6}
In this paper, we proposed a low-resolution image prior based network (LRIP-net) for the limited-angle reconstruction problems. We theoretically proved the low-resolution reconstruction problem has smaller condition number than the high-resolution problem, which performs stabler for the ill-posed problems. The constrained minimization problem was solved by the ADMM-based algorithm with each subproblem being approximated by the convolutional neural networks. Numerical experiments on various limited-angle CT reconstruction problems have successfully proved the advantages of our LRIP-net over the state-of-the-art learning methods.

We notice that the super-resolution technique has been used to improve the CT image resolution. The CT super-resolution method \cite{you2019ct} used a residual CNN-based network in the CycleGAN framework to recover high-resolution CT images. Although both methods involve the double resolution data, the train of thought is completely different. We use the low-resolution image as a prior for the limited-angle reconstruction problem, while they aimed to estimate high-resolution images from its low-resolution counterpart obtained by the FBP reconstruction.
Obviously, a possible future research direction is to use image super-resolution methods such as \cite{9007623}, \cite{SONG2021106193} to establish more effective connection between the low-resolution and high-resolution images for solving the ill-posed CT reconstruction problems. Another direction worth exploring is to develop effective methods to fuse the multi-scale information to obtain a priori image. Moreover, we also would like to investigate our method on other reconstruction problems such as digital breast tomosynthesis \cite{teuwen2021deep}, which is a limited-angle x-ray tomography technique using low-dose high-resolution projections.

\section*{Acknowledgement}
We would like to thank the anonymous referees for their valuable comments and helpful suggestions to improve the paper.




\begin{thebibliography}{10}
\providecommand{\url}[1]{#1}
\csname url@samestyle\endcsname
\providecommand{\newblock}{\relax}
\providecommand{\bibinfo}[2]{#2}
\providecommand{\BIBentrySTDinterwordspacing}{\spaceskip=0pt\relax}
\providecommand{\BIBentryALTinterwordstretchfactor}{4}
\providecommand{\BIBentryALTinterwordspacing}{\spaceskip=\fontdimen2\font plus
\BIBentryALTinterwordstretchfactor\fontdimen3\font minus
  \fontdimen4\font\relax}
\providecommand{\BIBforeignlanguage}[2]{{%
\expandafter\ifx\csname l@#1\endcsname\relax
\typeout{** WARNING: IEEEtran.bst: No hyphenation pattern has been}%
\typeout{** loaded for the language `#1'. Using the pattern for}%
\typeout{** the default language instead.}%
\else
\language=\csname l@#1\endcsname
\fi
#2}}
\providecommand{\BIBdecl}{\relax}
\BIBdecl

\bibitem{Frikel2013}
J.~Frikel, ``Sparse regularization in limited angle tomography,'' \emph{Applied
  and Computational Harmonic Analysis}, vol.~34, no.~1, pp. 117--141, Jan.
  2013.

\bibitem{He2019}
J.~He, Y.~Yang, Y.~Wang, D.~Zeng, Z.~Bian, H.~Zhang, J.~Sun, Z.~Xu, and J.~Ma,
  ``Optimizing a parameterized plug-and-play {ADMM} for iterative low-dose {CT}
  reconstruction,'' \emph{{IEEE} Transactions on Medical Imaging}, vol.~38,
  no.~2, pp. 371--382, Feb. 2019.

\bibitem{Goy2019}
A.~Goy, G.~Rughoobur, S.~Li, K.~Arthur, A.~I. Akinwande, and G.~Barbastathis,
  ``High-resolution limited-angle phase tomography of dense layered objects
  using deep neural networks,'' \emph{Proceedings of the National Academy of
  Sciences}, vol. 116, no.~40, pp. 19\,848--19\,856, Sep. 2019.

\bibitem{Candes2006}
E.~Candes, J.~Romberg, and T.~Tao, ``Robust uncertainty principles: exact
  signal reconstruction from highly incomplete frequency information,''
  \emph{{IEEE} Transactions on Information Theory}, vol.~52, no.~2, pp.
  489--509, Feb. 2006.

\bibitem{Lustig2008}
M.~Lustig, D.~Donoho, J.~Santos, and J.~Pauly, ``Compressed sensing {MRI},''
  \emph{{IEEE} Signal Processing Magazine}, vol.~25, no.~2, pp. 72--82, Mar.
  2008.

\bibitem{Chen2008}
G.-H. Chen, J.~Tang, and S.~Leng, ``Prior image constrained compressed sensing
  ({PICCS}): A method to accurately reconstruct dynamic {CT} images from highly
  undersampled projection data sets,'' \emph{Medical Physics}, vol.~35, no.~2,
  pp. 660--663, Jan. 2008.

\bibitem{2008Image}
E.~Y. Sidky and X.~Pan, ``Image reconstruction in circular cone-beam computed
  tomography by constrained, total-variation minimization,'' \emph{Physics in
  Medicine and Biology}, vol.~53, no.~17, p. 4777, 2008.

\bibitem{Ritschl2011}
L.~Ritschl, F.~Bergner, C.~Fleischmann, and M.~Kachelrie{\ss}, ``Improved total
  variation-based {CT} image reconstruction applied to clinical data,''
  \emph{Physics in Medicine and Biology}, vol.~56, no.~6, pp. 1545--1561, Feb.
  2011.

\bibitem{chen2013limited}
Z.~Chen, X.~Jin, L.~Li, and G.~Wang, ``A limited-angle {CT} reconstruction
  method based on anisotropic tv minimization,'' \emph{Physics in Medicine and
  Biology}, vol.~58, no.~7, p. 2119, 2013.

\bibitem{2014Edge}
A.~Cai, L.~Wang, H.~Zhang, B.~Yan, and J.~Li, ``Edge guided image
  reconstruction in linear scan {CT} by weighted alternating direction {TV}
  minimization,'' \emph{J Xray Sci Technol}, vol.~22, no.~3, pp. 335--349,
  2014.

\bibitem{Xu2019}
J.~Xu, Y.~Zhao, H.~Li, and P.~Zhang, ``An image reconstruction model
  regularized by edge-preserving diffusion and smoothing for limited-angle
  computed tomography,'' \emph{Inverse Problems}, vol.~35, no.~8, p. 085004,
  Jul. 2019.

\bibitem{2014Sparse}
S.~Niu, Y.~Gao, Z.~Bian, J.~Huang, W.~Chen, G.~Yu, Z.~Liang, and J.~Ma,
  ``Sparse-view x-ray {CT} reconstruction via total generalized variation
  regularization,'' \emph{Physics in Medicine and Biology}, vol.~59, no.~12, p.
  2997, 2014.

\bibitem{2017Euler}
H.~Zhang, L.~Wang, Y.~Duan, L.~Lei, G.~Hu, and B.~Yan, ``Euler's elastica
  strategy for limited-angle computed tomography image reconstruction,''
  \emph{IEEE Transactions on Nuclear Science}, vol.~64, no.~8, pp. 2395--2405,
  2017.

\bibitem{2017Block}
A.~Cai, L.~Li, Z.~Zheng, L.~Wang, and B.~Yan, ``Block-matching sparsity
  regularization-based image reconstruction for low-dose computed tomography,''
  \emph{Medical Physics}, vol.~45, no.~6, pp. 2439--2452, Apr. 2018.

\bibitem{wang2019guided}
J.~Wang, C.~Wang, Y.~Guo, W.~Yu, and L.~Zeng, ``Guided image filtering based
  limited-angle {CT} reconstruction algorithm using wavelet frame,'' \emph{IEEE
  Access}, vol.~7, pp. 99\,954--99\,963, 2019.

\bibitem{9084157}
M.~Xu, D.~Hu, F.~Luo, F.~Liu, S.~Wang, and W.~Wu, ``Limited-angle x-ray {CT}
  reconstruction using image gradient $\ell_0$-norm with dictionary learning,''
  \emph{IEEE Transactions on Radiation and Plasma Medical Sciences}, vol.~5,
  no.~1, pp. 78--87, 2021.

\bibitem{Wang}
C.~Wang, M.~Tao, J.~G. Nagy, and Y.~Lou, ``Limited-angle {CT} reconstruction
  via the {L1/L2} minimization,'' \emph{SIAM Journal on Imaging Sciences},
  vol.~14, no.~2, pp. 749--777, 2021.

\bibitem{pelt2013fast}
D.~M. Pelt and K.~J. Batenburg, ``Fast tomographic reconstruction from limited
  data using artificial neural networks,'' \emph{IEEE Transactions on Image
  Processing}, vol.~22, no.~12, pp. 5238--5251, 2013.

\bibitem{boublil2015spatially}
D.~Boublil, M.~Elad, J.~Shtok, and M.~Zibulevsky, ``Spatially-adaptive
  reconstruction in computed tomography using neural networks,'' \emph{IEEE
  Transactions on Medical Imaging}, vol.~34, no.~7, pp. 1474--1485, 2015.

\bibitem{kang2017deep}
E.~Kang, J.~Min, and J.~C. Ye, ``A deep convolutional neural network using
  directional wavelets for low-dose x-ray {CT} reconstruction,'' \emph{Medical
  Physics}, vol.~44, no.~10, pp. e360--e375, 2017.

\bibitem{gupta2018cnn}
H.~Gupta, K.~H. Jin, H.~Q. Nguyen, M.~T. McCann, and M.~Unser, ``{CNN}-based
  projected gradient descent for consistent {CT} image reconstruction,''
  \emph{IEEE Transactions on Medical Imaging}, vol.~37, no.~6, pp. 1440--1453,
  2018.

\bibitem{adler2018learned}
J.~Adler and O.~{\"O}ktem, ``Learned primal-dual reconstruction,'' \emph{IEEE
  Transactions on Medical Imaging}, vol.~37, no.~6, pp. 1322--1332, 2018.

\bibitem{chen2018learn}
H.~Chen, Y.~Zhang, Y.~Chen, J.~Zhang, W.~Zhang, H.~Sun, Y.~Lv, P.~Liao,
  J.~Zhou, and G.~Wang, ``{LEARN}: Learned experts¡¯ assessment-based
  reconstruction network for sparse-data {CT},'' \emph{IEEE Transactions on
  Medical Imaging}, vol.~37, no.~6, pp. 1333--1347, 2018.

\bibitem{han2018framing}
Y.~Han and J.~C. Ye, ``Framing {U}-net via deep convolutional framelets:
  Application to sparse-view {CT},'' \emph{IEEE Transactions on Medical
  Imaging}, vol.~37, no.~6, pp. 1418--1429, 2018.

\bibitem{zhang2019jsr}
H.~Zhang, B.~Dong, and B.~Liu, ``Jsr-net: a deep network for joint
  spatial-radon domain {CT} reconstruction from incomplete data,'' in
  \emph{ICASSP 2019-2019 IEEE International Conference on Acoustics, Speech and
  Signal Processing (ICASSP)}.\hskip 1em plus 0.5em minus 0.4em\relax IEEE,
  2019, pp. 3657--3661.

\bibitem{bubba2019learning}
T.~A. Bubba, G.~Kutyniok, M.~Lassas, M.~M{\"a}rz, W.~Samek, S.~Siltanen, and
  V.~Srinivasan, ``Learning the invisible: A hybrid deep learning-shearlet
  framework for limited angle computed tomography,'' \emph{Inverse Problems},
  vol.~35, no.~6, p. 064002, 2019.

\bibitem{arridge2019solving}
S.~Arridge, P.~Maass, O.~{\"O}ktem, and C.-B. Sch{\"o}nlieb, ``Solving inverse
  problems using data-driven models,'' \emph{Acta Numerica}, vol.~28, pp.
  1--174, 2019.

\bibitem{wurfl2018deep}
T.~W{\"u}rfl, M.~Hoffmann, V.~Christlein, K.~Breininger, Y.~Huang, M.~Unberath,
  and A.~K. Maier, ``Deep learning computed tomography: Learning
  projection-domain weights from image domain in limited angle problems,''
  \emph{IEEE Transactions on Medical Imaging}, vol.~37, no.~6, pp. 1454--1463,
  2018.

\bibitem{lin2019dudonet}
W.-A. Lin, H.~Liao, C.~Peng, X.~Sun, J.~Zhang, J.~Luo, R.~Chellappa, and S.~K.
  Zhou, ``Dudonet: Dual domain network for {CT} metal artifact reduction,'' in
  \emph{Proceedings of the IEEE/CVF Conference on Computer Vision and Pattern
  Recognition}, 2019, pp. 10\,512--10\,521.

\bibitem{baguer2020computed}
D.~O. Baguer, J.~Leuschner, and M.~Schmidt, ``Computed tomography
  reconstruction using deep image prior and learned reconstruction methods,''
  \emph{Inverse Problems}, vol.~36, no.~9, p. 094004, 2020.

\bibitem{ding2020low}
Q.~Ding, G.~Chen, X.~Zhang, Q.~Huang, H.~Ji, and H.~Gao, ``Low-dose {CT} with
  deep learning regularization via proximal forward--backward splitting,''
  \emph{Physics in Medicine and Biology}, vol.~65, no.~12, p. 125009, 2020.

\bibitem{cheng2020learned}
W.~Cheng, Y.~Wang, H.~Li, and Y.~Duan, ``Learned full-sampling reconstruction
  from incomplete data,'' \emph{IEEE Transactions on Computational Imaging},
  vol.~6, pp. 945--957, 2020.

\bibitem{Zang_2021_ICCV}
G.~Zang, R.~Idoughi, R.~Li, P.~Wonka, and W.~Heidrich, ``Intratomo:
  Self-supervised learning-based tomography via sinogram synthesis and
  prediction,'' in \emph{Proceedings of the IEEE/CVF International Conference
  on Computer Vision (ICCV)}, October 2021, pp. 1960--1970.

\bibitem{9496261}
D.~Hu, Y.~Zhang, J.~Liu, C.~Du, J.~Zhang, S.~Luo, G.~Quan, Q.~Liu, Y.~Chen, and
  L.~Luo, ``Special: Single-shot projection error correction integrated
  adversarial learning for limited-angle {CT},'' \emph{IEEE Transactions on
  Computational Imaging}, vol.~7, pp. 734--746, 2021.

\bibitem{doi:10.1137/20M1343075}
T.~A. Bubba, M.~Galinier, M.~Lassas, M.~Prato, L.~Ratti, and S.~Siltanen,
  ``Deep neural networks for inverse problems with pseudodifferential
  operators: An application to limited-angle tomography,'' \emph{SIAM Journal
  on Imaging Sciences}, vol.~14, no.~2, pp. 470--505, 2021.

\bibitem{hu2022dior}
D.~Hu, Y.~Zhang, J.~Liu, S.~Luo, and Y.~Chen, ``Dior: Deep iterative
  optimization-based residual-learning for limited-angle {CT} reconstruction,''
  \emph{IEEE Transactions on Medical Imaging}, vol.~41, no.~7, pp. 1778--1790,
  2022.

\bibitem{Jorgensen2013}
J.~S. Jorgensen, E.~Y. Sidky, and X.~Pan, ``Quantifying admissible
  undersampling for sparsity-exploiting iterative image reconstruction in x-ray
  {CT},'' \emph{{IEEE} Transactions on Medical Imaging}, vol.~32, no.~2, pp.
  460--473, Feb. 2013.

\bibitem{Jin2017}
K.~H. Jin, M.~T. McCann, E.~Froustey, and M.~Unser, ``Deep convolutional neural
  network for inverse problems in imaging,'' \emph{{IEEE} Transactions on Image
  Processing}, vol.~26, no.~9, pp. 4509--4522, Sep. 2017.

\bibitem{8363862}
H.~Yuan, J.~Jia, and Z.~Zhu, ``Sipid: A deep learning framework for sinogram
  interpolation and image denoising in low-dose {CT} reconstruction,'' in
  \emph{2018 IEEE 15th International Symposium on Biomedical Imaging (ISBI
  2018)}, 2018, pp. 1521--1524.

\bibitem{mccollough2017low}
McCollough \emph{et~al.}, ``Low-dose {CT} for the detection and classification
  of metastatic liver lesions: results of the 2016 low dose {CT} grand
  challenge,'' \emph{Medical Physics}, vol.~44, no.~10, pp. e339--e352, 2017.

\bibitem{Elbakri2002}
I.~Elbakri and J.~Fessler, ``Statistical image reconstruction for polyenergetic
  x-ray computed tomography,'' \emph{{IEEE} Transactions on Medical Imaging},
  vol.~21, no.~2, pp. 89--99, 2002.

\bibitem{you2019ct}
C.~You, G.~Li, Y.~Zhang, X.~Zhang, H.~Shan, M.~Li, S.~Ju, Z.~Zhao, Z.~Zhang,
  W.~Cong, and G.~Wang, ``{CT} super-resolution gan constrained by the
  identical, residual, and cycle learning ensemble (gan-circle),'' \emph{IEEE
  Transactions on Medical Imaging}, vol.~39, no.~1, pp. 188--203, 2019.

\bibitem{9007623}
F.~Fang, J.~Li, and T.~Zeng, ``Soft-edge assisted network for single image
  super-resolution,'' \emph{IEEE Transactions on Image Processing}, vol.~29,
  pp. 4656--4668, 2020.

\bibitem{SONG2021106193}
\BIBentryALTinterwordspacing
Z.~Song, X.~Zhao, Y.~Hui, and H.~Jiang, ``Progressive back-projection network
  for {COVID-CT} super-resolution,'' \emph{Computer Methods and Programs in
  Biomedicine}, vol. 208, p. 106193, 2021. [Online]. Available:
  \url{https://www.sciencedirect.com/science/article/pii/S0169260721002674}
\BIBentrySTDinterwordspacing

\bibitem{teuwen2021deep}
J.~Teuwen, N.~Moriakov, C.~Fedon, M.~Caballo, I.~Reiser, P.~Bakic,
  E.~Garc{\'\i}a, O.~Diaz, K.~Michielsen, and I.~Sechopoulos, ``Deep learning
  reconstruction of digital breast tomosynthesis images for accurate breast
  density and patient-specific radiation dose estimation,'' \emph{Medical Image
  Analysis}, vol.~71, p. 102061, 2021.

\end{thebibliography}
\end{document}